\pdfoutput=1
\documentclass[12pt]{iopart}
\usepackage{graphicx} 
\usepackage{dcolumn} 
\usepackage{array} 
\usepackage{bm} 
\usepackage{latexsym} 
\usepackage{epstopdf}
\usepackage{amsopn}
\newcommand{\sgn}{\operatorname{sgn}}
\begin{document} 

\title{Diffusive  motion of particles and dimers over anisotropic lattices} 

\author{Marcin Mi{\'n}kowski and Magdalena A Za{\l}uska--Kotur}
\address{Institute of Physics, Polish
Academy of Sciences, Al.~Lotnik{\'o}w 32/46, 02--668 Warsaw, Poland}
\ead{minkowski@ifpan.edu.pl, zalum@ifpan.edu.pl} 

\date{\today}

\begin{abstract}
Behavior of the mixture of   particles and dimers  moving with different jump rates at reconstructed surfaces is described. Collective diffusion coefficient is calculated by  the variational approach. Anisotropy of the  collective particle motion is analyzed  as a function of    jump rates and  local particle density. Analytic expressions are compared with the results of Monte Carlo simulations of diffusing particle and dimer mixture. Direction of  driven diffusive motion of the same system depends  on  the jump anisotropy and  on the value of driving force.  Driven motion results in the particle and dimer separation when the  directions of  their  easy  diffusion axes  differ. It is shown that in  such  case  trapping sites concentrated  at some surface areas    act as filters or barriers for particle and dimer mixtures.

\end{abstract}

\maketitle
\section{Introduction}
Collective  diffusion of particles   is a process  that is crucial for  various  surface phenomena\cite{ala-nissila}.  Emergence of different more or less regular  surface patterns is often controlled by  rate and anisotropy  of the surface diffusion\cite{growth,growth2,jeong,misbah}.  Diffusion of the adsorbed  gas is  described in many approaches by a number of   temperature activated single-particle jumps from one site to another. However,  motion  of dimers  or larger particle clusters  in  many systems is not just a simple combination of single-particle jumps.  Single-particle and dimer mobility was the subject of studies in  many systems\cite{manzi,bi,durr,ferrando,multi}. Dimer  diffusion quite often  appears to be higher and more important for the surface processes at least within some ranges of temperatures \cite{manzi,bi,durr,ferrando}.
Dimer jumps can also have   anisotropy different from  that of  single-particle jumps  \cite{montaleti1,montaleti2}. Below we study collective diffusion of a mixture of  particles and dimers jumping over the same, square lattice but with different   jump rates. Systems where single-particle jump rates were modified due to the presence of neighboring particles were studied in various context at lattices of different symmetries \cite{ala-nissila,montaleti1, montaleti2,lun,kjoll1,kjoll2,nieto, tarasenko}.  Similarly collective dynamics of many jumping dimers can be analyzed \cite{manzi}. At real surfaces dimers quite often occupy sites of different locations than single particles do\cite{montaleti1,montaleti2,lun}. In our model particles and dimers jump between the same   lattice sites. Dimers are represented by double occupancy of a site and are created when a particle jumps into a site already occupied by one particle. Once created dimer jumps as a single object or divides into two particles: one jumping out to a neighboring site and the second staying at the initial site. Single particles jump as single objects. An assumption that dimer occupies a single lattice site is a simplification that allows to derive diffusion coefficients for the system and  at the same part reproduces the main features  of the mixture dynamic. Both components diffuse over the same surface with jump rates of different anisotropy and there is some probability that dimer divides into two particles or that two neighboring particles change into dimer. 
 
Particle diffusion at most crystal surfaces is anisotropic. In particular it differs  in two main  directions along and across channels  emerging  at the surface of row  symmetry like W(110), W(112),W(322), Mo(111) or Ag(110) and many others \cite{antczak2,jalochowski,Ag(110),antczak,vat1,vat2,vat3}.   Some  surfaces, like Si(001)  reconstruct creating characteristic dimer row pattern and diffusion symmetry changes according to  the new surface structure    
\cite{bi,durr,antczak,kong,zhu}. The  reconstruction   implies  occurrence of different barrier heights for diffusion along different pathways. The anisotropy axis can vary depending whether it is a single particle or a dimer jumping at the same surface \cite{bi,durr,ferrando}. Mixture of particles and dimers forms in our model uniform  gas layer  at the surface.  Anisotropy at the surface is modeled by   the order of jump rates along lattice established independently for  each of the mixture components. Various realization of the anisotropy axis can be studied within the proposed model. We study system with different jump anisotropy  for single particles and  different for dimers. 
The consequences of such  kinetics  are studied for  the collective dynamics  and driven collective motion of the whole particle system.  
General expression for the diffusion coefficient   is derived for the lattice of given anisotropies for single particles and dimers. We also study the same system diffusing in the presence of an external driving force. We show how the dense cloud of  particles behaves when it diffuses freely and when it is driven along one of the lattice directions. Its geometrical shape changes  with decaying particle density in both cases. Diffusing gas of particles elongates along the main diffusion axis.  Main diffusion axes differ for particle and dimer motions, so the shape of diffusing cloud depends on its density and changes when  density  decreases.   Driven particle motion in general follows direction of the force, but it is turned towards  easy diffusion   axis. We analyze how the move of  driven cloud of particles and dimers depends on the driving force far above the limit of the  linear response. Next  we study motion of driven  gas across region with  trapping sites. We show  how the direction of  movement of diluted gas  of single particles changes when they come close to the properly oriented obstacle.  
\begin{figure}
\begin{center}
\includegraphics[width=7cm,angle=0]{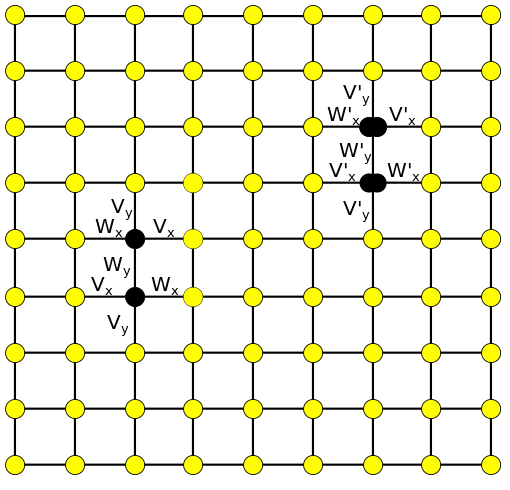}
\includegraphics[width=4cm,angle=0]{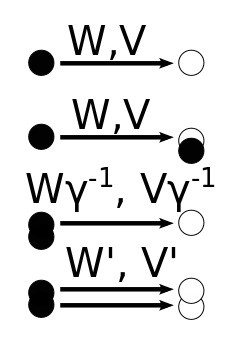}
\caption{Schema of the lattice with jump rates of  single particle at the left side of the lattice and of  dimer at the right side of the lattice.
 Particles and dimers jump over the same lattice sites. Two particles occupying the same site create one dimer. It can break into two particles with the probability $\gamma^{-1}$. At right side four types of jumps are plotted from top down: single particle to the empty site,  single particle to the occupied site, one single particle from dimer into the empty, or occupied by single particle site  and dimer into the empty site. }
\label{lattice}
\end{center}
\end{figure}
\section{The model} 
\subsection{Equilibrium density for particles, dimers and their motion over the lattice}
A mixture of single particles and dimers occupying the same lattice sites is studied below. In principle dimers  occupy  different positions at the surface  \cite{bi,durr,ferrando}. Below, the simplified model is described, where we assume that  two particles that build dimers sit at the same  lattice site. No other objects are allowed and the maximal site occupation is by one dimer, what means two particle per lattice site. The equilibrium densities  of single particles $\theta_1$ and dimers $\theta_2$ are determined by the the chemical activity $z$ and the energy factor describing  dimer creation   $\gamma=\exp(\beta J)$ where $\beta=1/{k_B T}$ is temperature $T$ factor and $J$ is the energy of dimer bond. Site occupation can be expressed as
\begin{eqnarray}
\theta_1=\frac{z}{1+z+z^2 \gamma} \nonumber  \\
\theta_2=\frac{z^2 \gamma}{1+z+z^2 \gamma}  \label{static}\\
\theta=\theta_1+2 \theta_2  \nonumber
\end{eqnarray}
where $\theta_1$ means local density of single particles, $\theta_2$ local density of dimers and
$\theta$ is mean particle density  at the  site.  Note that $\theta$ changes from 0 to 2, whereas  in the case when no dimers are formed in the system the maximal value of $\theta$ is 1.

We consider collective diffusion of a two-dimensional lattice gas. Jumps  of single particles from site to site are given by  four  different constants:  $W_x$, $V_x$  along axis $x$  and $W_y$, $V_y$ along  axis $y$.  They are arranged alternately, so as an effect   lattice contains two types of sites, with different pattern of jump rates away  from each of them.  The main anisotropy axis can be oriented   at any angle to the axes $x$ and $y$, depending on the values of  jump rates. The heights of the potential barriers between sites are different for each of four types of jumps, while the  local equilibrium  energy is the  same at  each site. Besides single-particle jumps we also take into account creation of  dimers and their tendency to move in their own way. They jump over the lattice with jump rates  given by: $ V\rq{}_x, V\rq{}_y, W\rq{}_x, W\rq{}_y $. There is a probability $\gamma^{-1}$ that dimer separates into individual particles and then each of them jumps as a single particle. The lattice  is illustrated at the left side of Fig 1, where pattern of single-particle and dimer jumps from equivalent lattice sites is shown. Possible jumps in the direction $x$ are enumerated at the right side of Fig. 1.

\subsection{Collective diffusion coefficient}
 We study the  collective  motion of  the system, i.e. many particles are  treated as one assembly.
 Time evolution of this system is governed by the set of Markovian master rate equations for the probabilities $P(\{n\},t)$ that a microscopic microstate $\{n\}$ of a lattice gas occurs at time $t$
\begin{eqnarray}
  \label{eq:1}
  \frac{d}{dt} P(\{n\},t) =
  \sum_{\{n'\}} [{\cal W}(\{n\},\{n'\})
  P(\{n'\},t)-{\cal W}(\{n'\},\{n\}) P(\{n\},t) ].
\end{eqnarray}
$\{n\}$ is understood as a set of variables specifying which
particular sites in the lattice are occupied and which are not. Double occupancy is forbidden i.e. $n_i=0$ for an empty site $n_i=1$ for a particle or $n_i=2$ for a dimer and transition rates ${\cal W}(\{n\},\{n'\})$ depend on the local potential energy landscape. 
 For a single particle ${\cal W} =V_x, V_y, W_x, W_y $  depending on the initial site and the jump direction.  Other set  of rates ${\cal W} =V\rq{}_x, V\rq{}_y, W\rq{}_x, W\rq{}_y $ defines  dimer kinetics.  Collective diffusion coefficient describes system diffusion and will be calculated using  the  variational approach. The variational approach to collective diffusion on  such a lattice was first proposed in \cite{GZ-K} and then applied in \cite{Z-KG1,BZ-KG1,KZ-K,min} for various systems.  It relates the collective diffusion coefficient, which in general is given by a $ 2 \times 2$ matrix $\hat{D}$ accounting for anisotropy, to the lowest eigenvalue of the rate matrix $\hat M(\vec{k})$. 
Matrix elements of  $\hat M$ are equal to ${\cal W}(\{n\},\{n\rq{}\})$ for almost all particle and dimer jumps  which must be additionally   multiplied by $exp(\pm i \vec{k}\vec{L})$ when configurations  $\{n\}$ and $\{n\rq{}\}$ differ by position of the  arbitrary chosen reference particle. Parameter $\vec{L}$ is a vector with components being  the system sizes $L_x, L_y$ and all results are taken as the infinite limit of these parameters and of the number $N$ of particles in the system. The lowest eigenvalue of the rate matrix is given by
\begin{equation}
\lambda_D^{var}\left(\vec{k}\right)\equiv\frac{\tilde{\phi}\left(\vec{k}\right)\cdot
\left[-\hat{M}\left(\vec{k}\right)\right]\cdot\phi\left(\vec{k}\right)}{\tilde{\phi}\left(\vec{k}\right)\cdot\phi \left(\vec{k}\right)} \approx_{\vec{|k|} -\rightarrow 0} -\vec{k}\hat{D}\vec{k}\label{eigenvalue}
\end{equation}
$ \phi\left(\vec{k}\right) $ is a trial eigenvector in this variational approach.The rate matrix is non Hermitian requiring to distinguish between right and left (having tilde  $ \tilde{} $ above it) eigenvectors. Rate matrix $ \hat{M }$, eigenvalue $\lambda$ and eigenvector $\phi$ are defined  in the Fourier space of $\vec{k}$ wavevectors.  Trial eigenvector for the studied systems is assumed  to be \cite{KZ-K}
\begin{equation}
\phi\left(\vec{k}\right)=\sum_{j=0}^{N-1}e^{i\left(k_xx_j+k_yy_j+k_x\delta_j^x+k_y\delta_j^y\right)}
\end{equation}
where the lattice position of the j-th particle is given by the vector $\left(x_j,y_j\right)$.  The variational parameters $\delta_i^x$ and $\delta_i^y$ are to be found in such a way that $\lambda_D^{var}$ has the smallest possible value. N is the total number of the particles in the system which tends to infinity together with $L_x$ and $L_y$.
$\hat{D}$ can be also written as the ratio
\begin{equation}
\vec{k}\hat{D}\vec{k}=\frac{{\cal M} \left(\vec{k}\right)}{{\cal N}\left(\vec{k}\right)}
\label{D}
\end{equation}
Denominator $\cal N$ is called static factor, it depends only on the equilibrium properties and can be calculated   as
\begin{equation}
{\cal N}= z \frac {d \theta}{d z},
\label{N}
\end{equation}
whereas the numerator $ \cal M $ contains all details of the particle and dimer kinetics.
\begin{figure}
\begin{center}
\includegraphics[width=6cm]{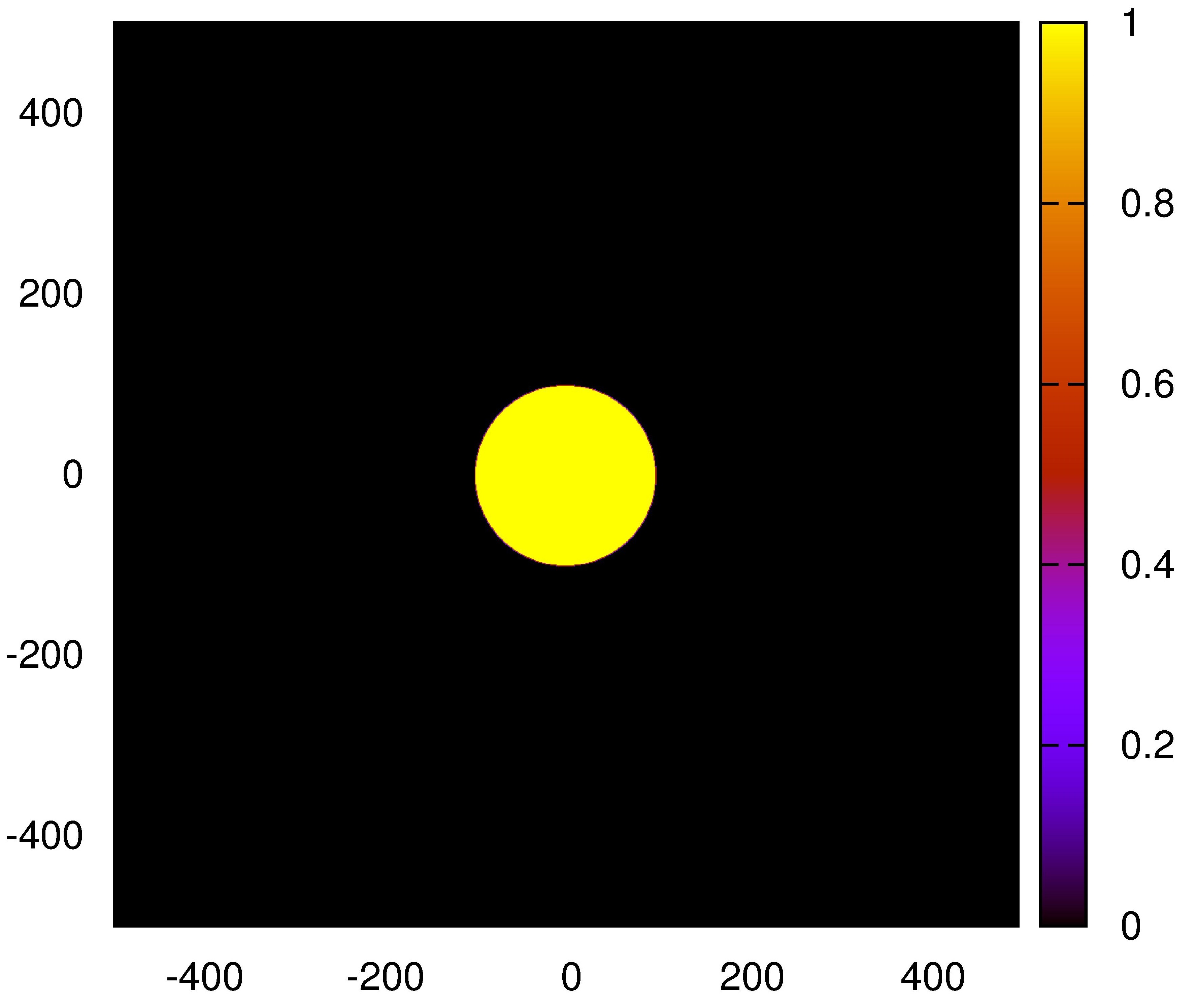}
\caption{Initial particle configuration}
\label{initial}
\end{center}
\end{figure}
\begin{figure}
\begin{center}
\includegraphics[width=6cm]{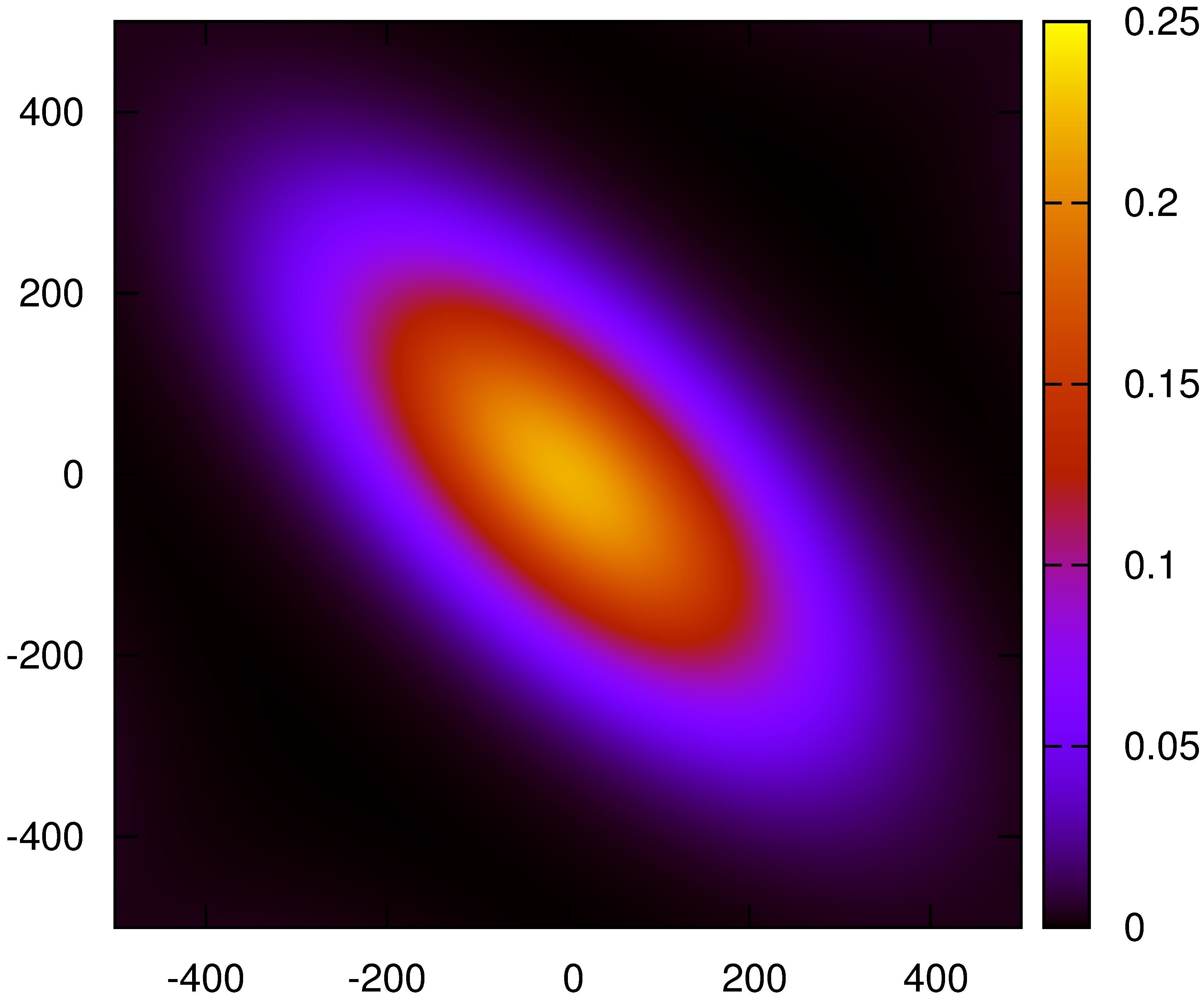}
\includegraphics[width=6cm]{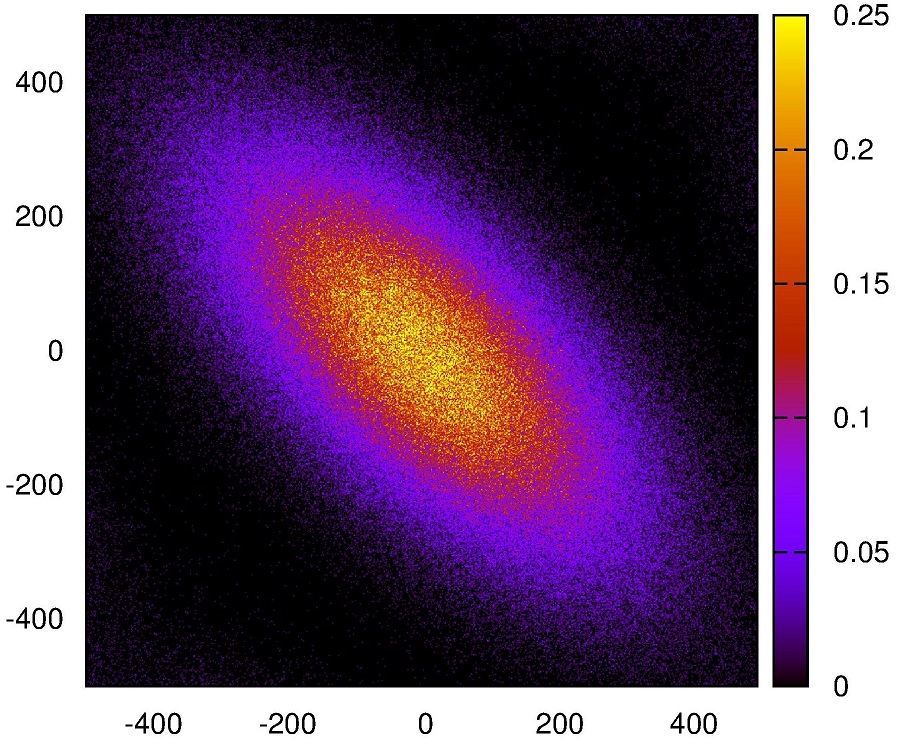}
\caption{Density plots of the diffusion of a non-interacting gas after t=19200 $ \tau$. $\tau$ is time unit. The left hand side plot shows the results of the analytical results and the right hand side plot shows the Monte Carlo simulations. The jump rates are $W_x=3 a^2/\tau$, $W_y=2 a^2/\tau$, $V_x=0.1 a^2/\tau$ and $V_y=0.1 a^2/\tau$.}
\label{no_interactions}
\end{center}
\end{figure}
\section{Anisotropic collective diffusion}
\subsection{Collective diffusion of single-particle system}
Let us first analyze system with single particles only. There are no dimers jumping over the lattice,   $\theta_2=0$.
In such case all particles at each site have the same energy  and denominator ${\cal N}\left(\vec{k}\right)$   reduces to 
\begin{equation}
{\cal  N}\left(\vec{k}\right)=N \theta\left(1-\theta\right) \label{NormDen}
\end{equation}
where $\theta = \theta_1$ is the particle density and $N$ is the number of lattice sites. To calculate the  numerator ${\cal M}\left(\vec{k}\right)$ we note that due to the system symmetry every second site is different. We have $\delta_W^x$ and $\delta_W^y$
for one type of sites, and $\delta_V^x$ and $\delta_V^y$ for the second. In such a case two new variational parameters $\delta^x=\delta_V^x-\delta_W^x$ and $\delta^y=\delta_V^y-\delta_W^y$ can be used, see Ref \cite{BZ-KG1} for details. After taking into account all the jumps with their probabilities and phase changes we find that the eigenvalue's numerator is
\begin{eqnarray}
{\cal M}&=&\frac{N}{2}\theta\left(1-\theta\right)\{W_x\left(k_x+k_x\delta^x-k_y\delta^y\right)^2\nonumber\\
&&+W_y\left(k_y+k_x\delta^x - k_y\delta^y\right)^2+V_x\left(k_x-k_x\delta^x+k_y\delta^y\right)^2\nonumber\\
&&+V_y\left(k_y-k_x\delta^x+k_y\delta^y\right)^2\}
\end{eqnarray}
and the values of $\delta^x$ and $\delta^y$, obtained after minimizing ${\cal M}$ with respect to them are
\begin{eqnarray}
\delta^x=\frac{V_x-W_x}{\left(W_x+V_y+V_x+W_y\right)}\\
\delta^y=\frac{V_y-W_y}{\left(W_x+V_y+V_x+W_y\right)}\nonumber
\end{eqnarray}
This  according to (\ref{D}) leads to
\begin{equation}
\hat{D}=\left(\begin{array}{cc}
\frac{(V_x+W_x)(V_y+W_y)+4V_xW_x}{2(W_x+V_x+W_y+V_y)} & \frac{-(V_x-W_x)(V_y-W_y)}{4(W_x+V_x+W_y+V_y)} \\
\frac{-(V_x-W_x)(V_y-W_y)}{4(W_x+V_x+W_y+V_y)} & \frac{(V_x+W_x)(V_y+W_y)+4V_y W_y}{2(W_x+V_x+W_y+V_y)}
\end{array}\right)\label{diffusion_matrix}
\end{equation}
Having the  analytic form of the diffusion coefficient   we can calculate the  particle density at  time $t$  as 
\begin{equation}
\rho \left(\vec{r},t\right)=\int\limits_{\Omega}\frac{1}{4\pi t\sqrt{\det{\hat{D}}}}\exp{\left[-\frac{\left(\vec{r}-\vec{r\rq{}}\right) \hat{D}^{-1}\left(\vec{r}-\vec{r\rq{}}\right)}{4t}\right]}d^2{r\rq{}}\label{int}
\end{equation}
for arbitrary initial particle configuration $\Omega$. 
 We have calculated  $\left(\ref{int}\right)$  numerically starting from the initial configuration shown in Fig \ref{initial}. In this calculation all the particles reside  at $t=0$ in a circle of the radius $r_0=100$ 
  \begin{eqnarray}
  \theta(\vec{r\rq{}},t=0) = \left\{ \right. 
 & \quad  \ \  \ 1 \ \ \ \ \  {\rm for}  \ \ \ \ \vec{r\rq{}}  \in \Omega  \nonumber \\
     & \quad \ \ \  0  \ \ \ \ {\rm for } \ \ \ \ \vec{r\rq{}}  \notin \Omega   \nonumber
 \end{eqnarray} 
 An exemplary set of parameters $W_x=3 a^2/\tau $, $W_y=2 a^2/\tau $, $V_x=0.1 a^2/ \tau$ and $V_y=0.1 a^2/ \tau$ has  been chosen for calculations.  Resulting particle density profile after some integration time is plotted in Fig \ref{no_interactions} on left side.
Kinetic Monte Carlo simulations have been also done for  the same system  and density profile is plotted at right hand side of Fig. \ref{no_interactions}.  In the simulations results were averaged over 20 samples.
It can be seen that density profiles obtained after integration and from MC data are almost identical. To compare them more precisely we  drew profiles of density for certain values of $y$. Monte Carlo results after coarse graining  over $6 \times 20 $ sites lie exactly at the analytical curve for all three cuts.
\begin{figure}
\begin{center}
\includegraphics[width=8cm,angle=0]{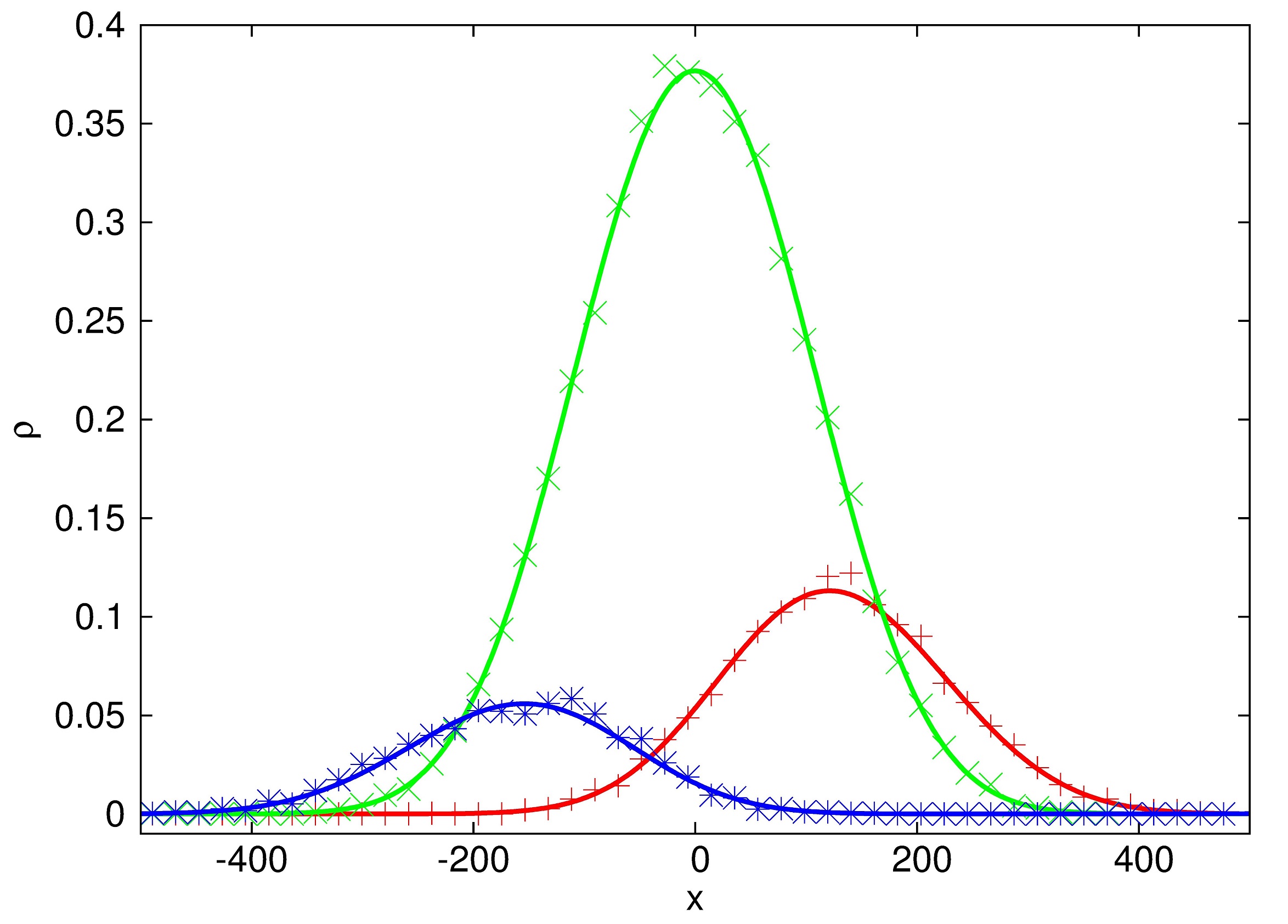}
\caption{Profiles of density for $y=-200$ (red line); $0$ (green line) and $250$ (blue line). Lines are plotted for the analytical results and points show averaged results of the Monte Carlo simulations from  Fig 1 at $t=19200 \tau$.}
\label{profile}
\end{center}
\end{figure}

\begin{figure}
\begin{center}
\includegraphics[width=8cm,angle=0]{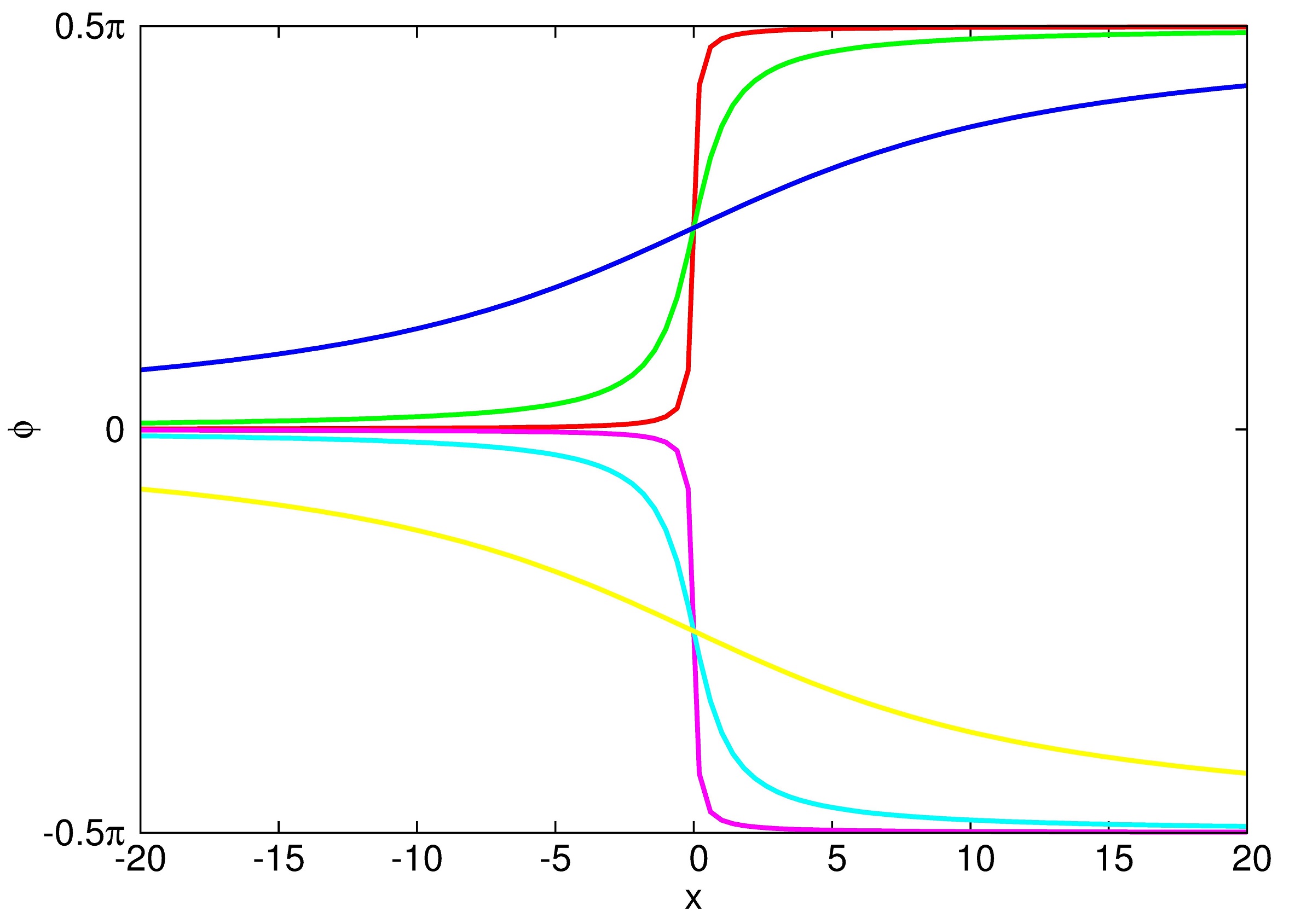}
\caption{Dependence of the diffusion angle on the variable $\alpha=2\left(V_yW_y-V_xW_x\right)$. Curves in the upper part of the plot are for positive values of $\beta=\left(V_x-W_x\right)\left(V_y-W_y\right)$ and those in the lower part are for negative values. Beginning from the steepest curve $\beta=\pm 0.1;\pm1;\pm10$.}
\label{angle}
\end{center}
\end{figure}

Both Figures \ref{no_interactions} and \ref{profile} illustrate evident anisotropy of the particle cloud. We can calculate eigenvalues of the diffusion matrix (\ref{diffusion_matrix})
\begin{eqnarray}
D_{\pm} &=&  \frac{1}{2}\Sigma_1 \pm \sqrt{ \frac{1}{4}{\Sigma_1}^2 - \Sigma_2 } \\
\Sigma_1 &=&  \frac{(V_x+V_y)(W_x+W_y)+(W_x+V_y)(V_x+W_y)} {V_x+W_x+V_y+W_y}
\nonumber  \\
\Sigma_2&=&\frac{ V_xV_yW_x+V_xV_yW_y+V_xW_xW_y+V_yW_xW_y}{V_x+W_x+V_y+W_y} \nonumber
\end{eqnarray}
$D_+$ and $D_-$ represent diffusion coefficients along two main axes of diffusion, the faster and the slower one respectively and the first of them denotes faster, the second slower direction. The angle of the diffusion anisotropy i.e. the angle between the direction of faster diffusion and the  $x$ axis is  
\begin{eqnarray}
\phi=\arctan\left(\frac{\alpha}{\beta}-\sgn({\beta})\sqrt{\left(\frac{\alpha}{\beta}\right)^2+1}\right)\\
\alpha=2(V_yW_y-V_xW_x) \nonumber \\
\beta=(V_x-W_x)(V_y-W_y) \nonumber
\end{eqnarray}
The dependence of the diffusion angle on the combination of transition rates denoted above as $\alpha$ is plotted in Fig \ref{angle}. Both cases for $\beta >0$ and $\beta <0$ are presented  in Fig \ref{angle}. There are two branches of solutions, one for positive and one for negative $\beta$. It can be seen that for $\alpha<0$ diffusion has its main direction  close to the $x$ axis above or below it depending on the sign of parameter $ \beta$. When $\alpha> 0$ diffusion happens preferentially along $y$ axis.

\subsection{Collective diffusion of particle and dimer mixture}
Let us now analyze the situation when diffusion of dimers  differs from that of  single particles. Dimer in our model means that two particles occupy the same site  and they  jump to the neighboring sites with  rates $W_x\rq{}$, $W_y\rq{}$,$V_x\rq{}$, $V_y\rq{}$. It is also possible that only one particle  jumps out from the dimer with  rates  the same as that for single particles $W_x$, $W_y$,$V_x$, $V_y$. The probability of such event is given by $\gamma^{-1}$.
Consequently  diffusion coefficient depends on the particle  density. Denominator  (\ref{N}) for this system can be calculated on using  Eqs (\ref{static}) as
\begin{equation}
{\cal N}= N [\theta (2-\theta)-\theta_1].
\label{den}
\end{equation}
 Denominator (\ref{den}) is zero for limit values of $\theta=0,2$. 
The numerator depends on three types of variational parameters: $\delta_S$,$\delta_D$ and $\delta_C$ which refer to  jumps of single particles, dimers and  jumps of single particles from one dimer to the other, respectively. Each of these phases occurs in  $x$ or $y$ direction.  Taking into account all possible transitions we have
\begin{eqnarray}
{\cal M}=\frac{N}{2}\{ W_{x}(1-\theta_1-\theta_2)\theta_1\left[k_x(1+\delta_S^x)-k_y\delta_S^y\right]^2 + 2 W_x {\theta_1}^2\left[k_x(1+\delta_C^x)- k_y\delta_C^y\right]^2 \nonumber \\ W_{x}\theta_1^3\left[k_x(1-\delta_S^x+\delta_D^x)+k_y(\delta_S^y -\delta_D^y)\right]^2 +  W\rq{}_x {\theta_2}(1-\theta_1-\theta_2) \left[k_x(2+\delta_D^x)-k_y\delta_D^y\right]^2 + \nonumber\\
 V_{x}(1-\theta_1-\theta_2)\theta_1\left[k_x(1-\delta_S^x)+k_y\delta_S^y\right]^2 + 2 V_x {\theta_1}^2\left[k_x(1-\delta_C^x)+k_y\delta_C^y\right]^2 \nonumber \\ V_{x}\theta_1^3\left(k_x(1+\delta_S^x-\delta_D^x)-k_y(\delta_S^y -\delta_D^y)\right]^2 +  V\rq{}_x {\theta_2}(1-\theta) \left[k_x(2-\delta_D^x)+k_y\delta_D^y\right]^2 + \nonumber \\
 W_{y}(1-\theta_1-\theta_2)\theta_1\left[k_y(1+\delta_S^y)-k_x\delta_S^x\right]^2 + 2 W_y {\theta_1}^2\left[k_y(1+\delta_C^y)- \delta_C^x\right]^2 \nonumber \\ W_{y}\theta_1^3\left[k_y(1-\delta_S^y+k_y\delta_D^y)+k_x(\delta_S^x -\delta_D^x)\right]^2 +  W\rq{}_y {\theta_2}(1-\theta_1-\theta_2) \left[k_y(2+\delta_D^y)-k_y\delta_D^y\right]^2 + \nonumber\\
 V_{y}(1-\theta_1-\theta_2)\theta_1\left[k_y(1-\delta_S^y)+k_x\delta_S^x\right]^2 + 2 V_y {\theta_1}^2\left[k_y(1-\delta_C^y)+k_x\delta_C^x\right]^2+ \nonumber \\ V_{y}\theta_1^3\left(k_y(1+\delta_S^y-\delta_D^y)-k_x(\delta_S^x -\delta_D^x)\right]^2 +  V\rq{}_y {\theta_2}(1-\theta_1-\theta_2) \left[k_y(2-\delta_D^y)+k_x\delta_D^x\right]^2 \}
\label{int_diff}
\end{eqnarray}
and the  phases that minimize this expression are
\begin{eqnarray}
\delta_C = \frac{V_x-W_x}{W_x+W_y+V_x+V_y} \nonumber \\
\delta_D=\frac{2(\tilde{V}_x-\tilde{W}_x)}{\tilde{W}_x+\tilde{W}_y+\tilde{V}_x+\tilde{V}_y} \nonumber \\
\delta_S=\frac{1-z^2}{1+z^2} \delta^C +2 \frac{z^2}{1+z^2} \delta^D
\end{eqnarray}
where
\begin{eqnarray}
\tilde{W}_{x,y}=z\gamma^{-1} W_{x,y} +(1+z^2) W\rq{}_{x,y} \nonumber \\
\tilde{V}_{x,y}=z\gamma^{-1} V_{x,y} +(1+z^2) V\rq{}_{x,y} 
\label{tilde}
\end{eqnarray}
The diffusion coefficient matrix looks now as follows
\begin{eqnarray}
\hat{D}_{dip}= \frac{1}{\theta(2-\theta)-\theta_1}\left[\frac{\theta_1(1-\theta_1-\theta_2)(1-z^2)^2}{1+z^2}+2 {\theta_1}^2\right]\hat{D} \nonumber \\+ \frac{1}{\theta(2-\theta)-\theta_1}\frac{4 \theta_2 (1-\theta_1-\theta_2)}{1+z^2}\left(\begin{array}{cc}{\tilde D}_{xx}
 &{\tilde D}_{xy}  \\ {\tilde D}_{yx}
 &{\tilde D}_{yy}
\end{array}\right)\label{diffusion_matrix_dip}
\end{eqnarray}
where $\hat D$ is  diffusion matrix given by Eq. (\ref{diffusion_matrix}) and
\begin{eqnarray}
{\tilde D}_{xx}=\frac{(\tilde V_x+\tilde W_x)(\tilde V_y+\tilde W_y)+4\tilde V_x  \tilde W_x}{2(\tilde W_x+\tilde V_x+\tilde W_y+\tilde V_y)}\\
\tilde D_{xy}= \tilde D_{y,x}=\frac{-( \tilde V_x-\tilde W_x)(\tilde V_y-\tilde W_y)}{4(\tilde W_x+\tilde V_x+\tilde W_y+\tilde V_y)} \\
{\tilde D}_{yy}=\frac{(\tilde V_x+\tilde W_x)(\tilde V_y+\tilde W_y)+4\tilde V_y  \tilde W_y}{2(\tilde W_x+\tilde V_x+\tilde W_y+\tilde V_y)}
\end{eqnarray}
are expressed by coefficients (\ref{tilde}). For single particles diffusion we come back to (\ref{diffusion_matrix}) and dimer diffusion when  $\theta_1=0$ is given by dimer jump rates $ W\rq{}_{x,y},V\rq{}_{x,y}$ only. However, in totally occupied system  $\theta=2$ diffusion does not happen by dimer jumps, but  by the single particles that separate from one dimer and join the next one.
We can now 
analyze diffusion of the dimer and single particle mixture. It is assumed that jump rates of single particles are the same as in Fig. (\ref{no_interactions})  and jump rates for dimers are as follows: $W_y\rq{}=0. 2 a^2/\tau $, $V_y\rq{}=5 a^2/ \tau$ and $W_x\rq{}=W_x$ $V_x\rq{}=V_x$. With such choice of the jump rates the  main axes of the  single particle and dimer diffusion  are  along two diagonals of the lattice.
In Fig. 6 the dependence of the matrix $\hat D_{dip}$ elements are plotted as a function of the system density $\theta$. Their limit value for $\theta=0$ is the same  for  single particle diffusion (\ref{diffusion_matrix}) and the second limit $\theta=2$  is given by the same values, but divided by the parameter $\gamma$. Dimer diffusion can be seen for intermediate densities from $0.1-0.8$, where the  parameter $D^{dip}_{xy}$  changes  from the negative to the positive value, thus rotating diffusion direction from one diagonal to the other.
 
 Diffusion of an initially circular spot of density $\theta=2$ given by the diffusion coefficient (\ref{int_diff}) was calculated and is shown in Fig \ref{interactions} on the left side of the plot.
\begin{figure}
\begin{center}
\includegraphics[width=8cm,angle=0]{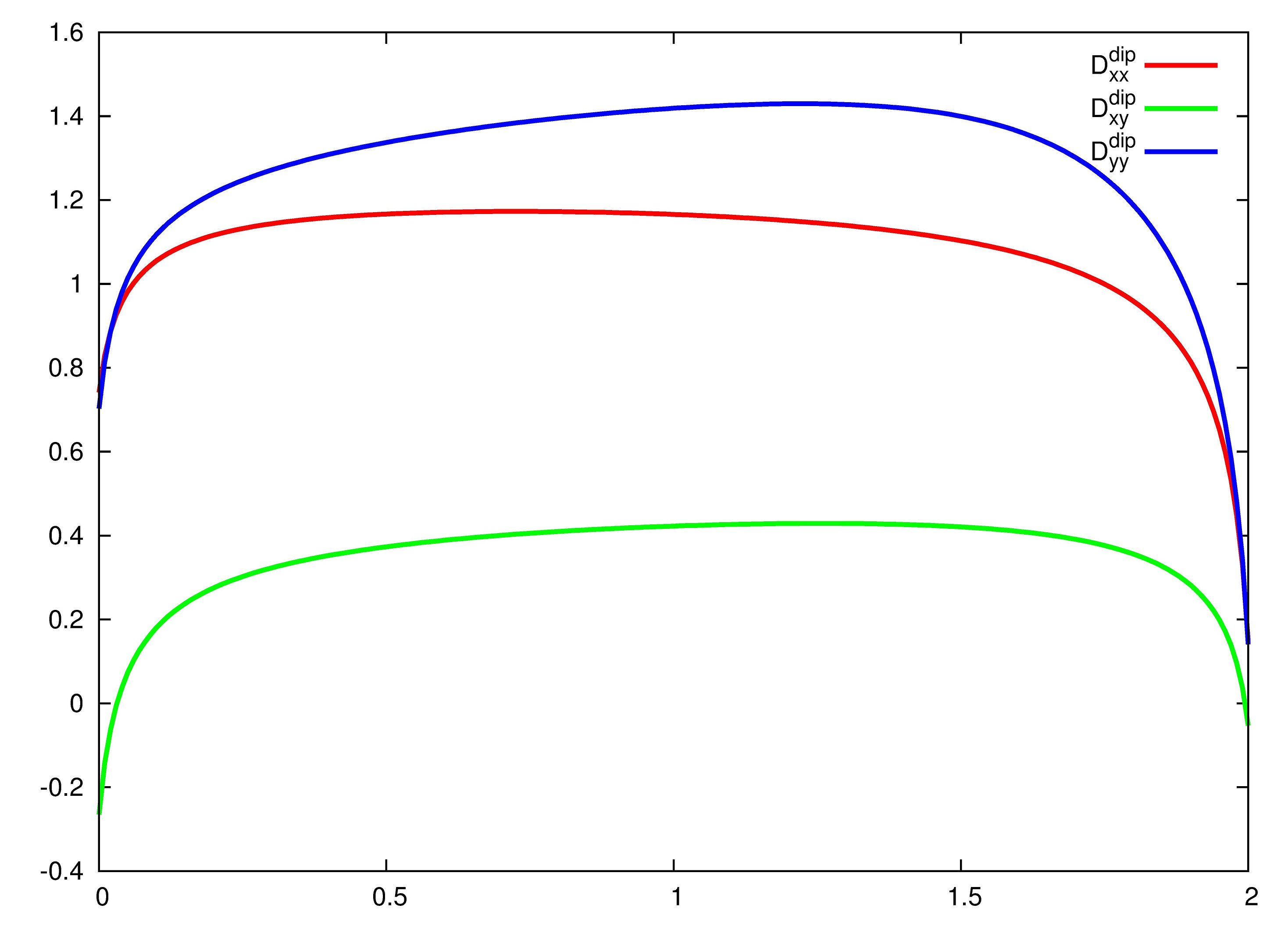}
\caption{Dependence of the elements of  diffusion  coefficient  matrix (\ref{diffusion_matrix_dip}) on the total system density $\theta$.$W_x=3 a^2/\tau $, $W_y=2 a^2/\tau $, $V_x=0.1 a^2/ \tau$, $V_y=0.1 a^2/ \tau$$W_y\rq{}=0. 2 a^2/\tau $, $V_y\rq{}=5 a^2/ \tau$ and $W_x\rq{}=W_x$ $V_x\rq{}=V_x$.}
\label{D_dip}
\end{center}
\end{figure}
\begin{figure}
\begin{center}
\includegraphics[width=5cm]{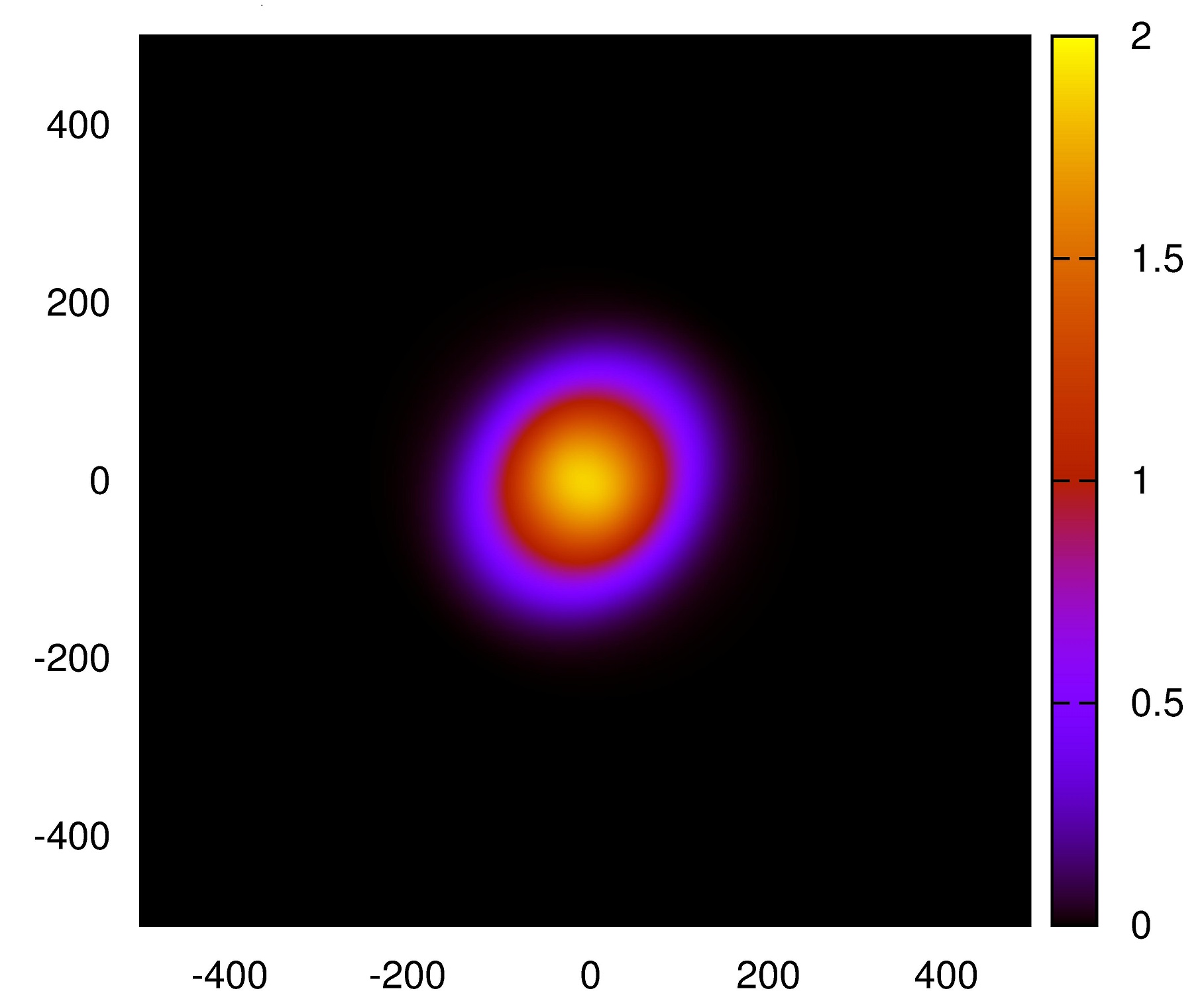}
\includegraphics[width=5cm]{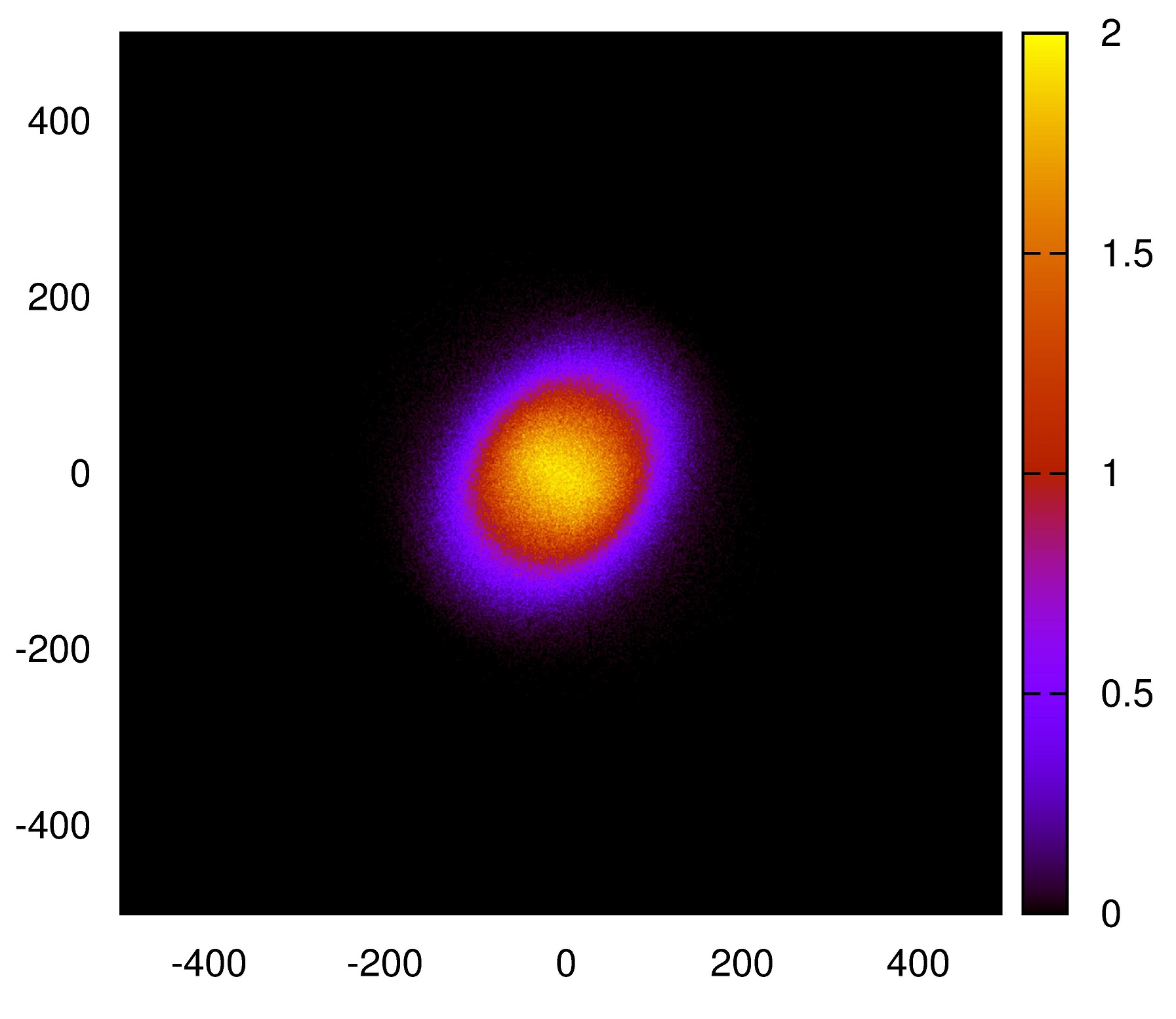}
\end{center}
\begin{center}
\includegraphics[width=5cm,angle=0]{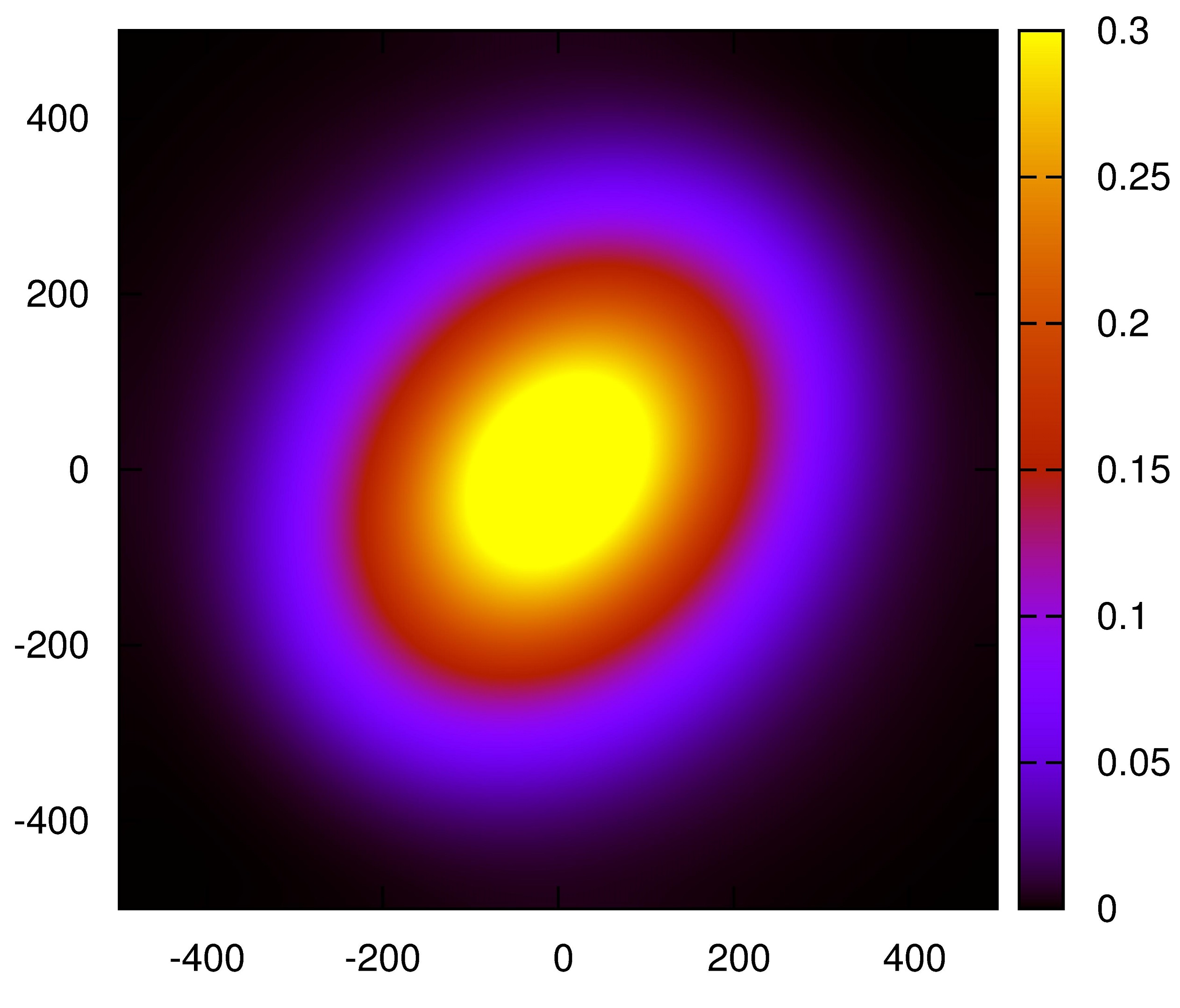}
\includegraphics[width=5cm]{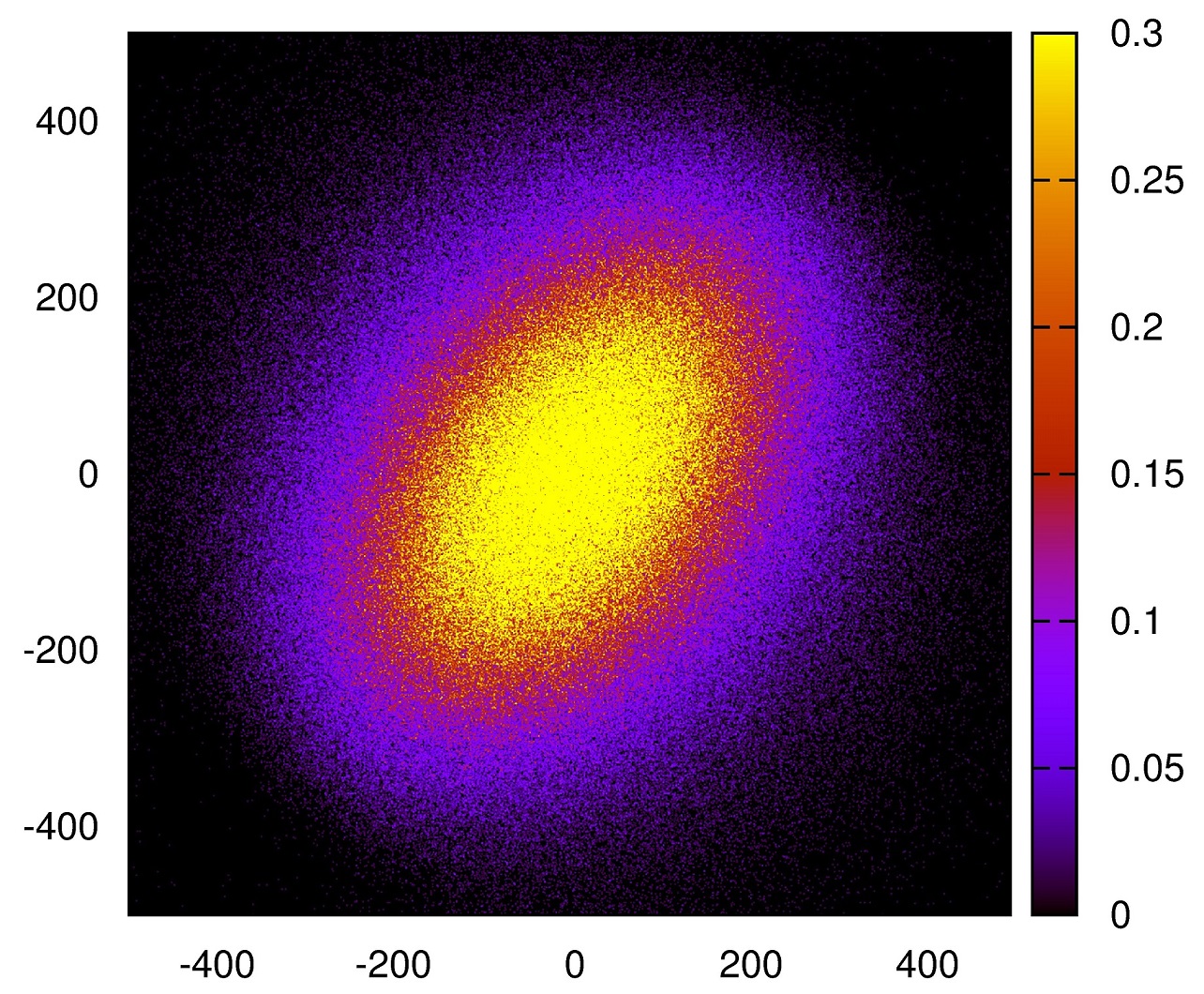}
\end{center}
\caption{Density plots of the diffusion of an interacting gas after $ t=700 \tau$ (upper plot) and $ t=13000 \tau$ (lower plot). Analytic results at left hand side and MC simulations at right hand side. }
\label{interactions}
\end{figure}
It can be seen that the core of the particle cloud, where the density is larger, diffuses along one diagonal, whereas single particles in the rarefied edge of the cloud move  along the other diagonal.  
 In Fig.\ref{interactions} on the right side  Monte Carlo density plots can be compared with integrated density profiles for the same times. We can see that both profiles are similar. 
Diffusion matrix (\ref{diffusion_matrix_dip}) describes equilibrium diffusion of the particle cloud over anisotropic lattice for the mixture of dimers and single particles diffusing in different way. We can see that dense regions  behave in other way than the rare ones. The overall density profile  of the mixture depends on the relative    dimer and   single particle mobility.
\begin{figure}
\begin{center}
\includegraphics[width=8cm,angle=0]{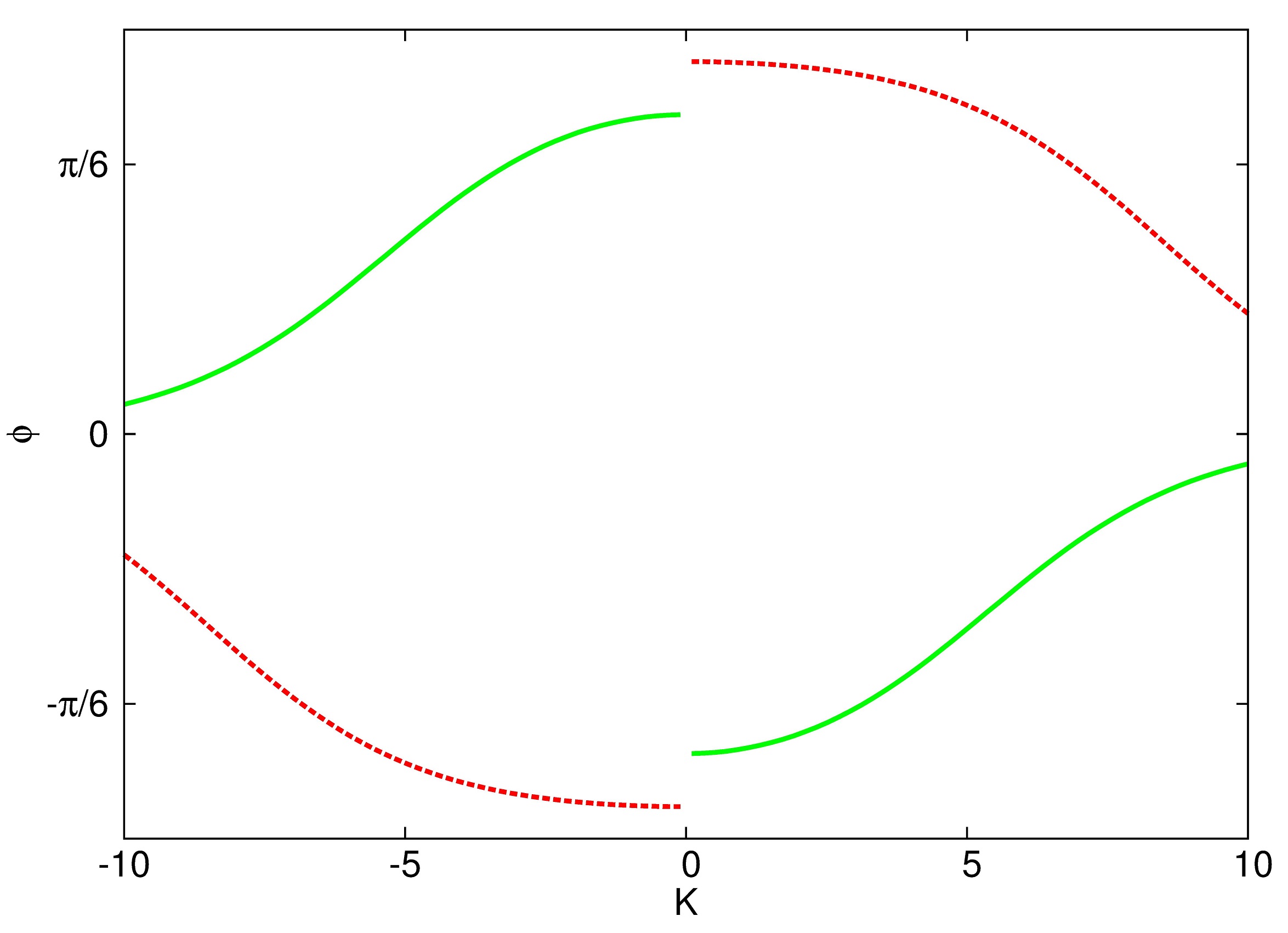}
\end{center}
\caption{Dependence of the angle of net current  with respect to the x-axis on the external driving force (\ref{angle_current}). The angle  calculated  for free particles is plotted by the solid (green) curve and for dimers is plotted  by the dashed (red).  Transition rates for single particles and for dimers are the same as those used  in Fig \ref{interactions}}
\label{angle_bias}
\end{figure}
\begin{figure}
\begin{center}
\includegraphics[width=4.2cm]{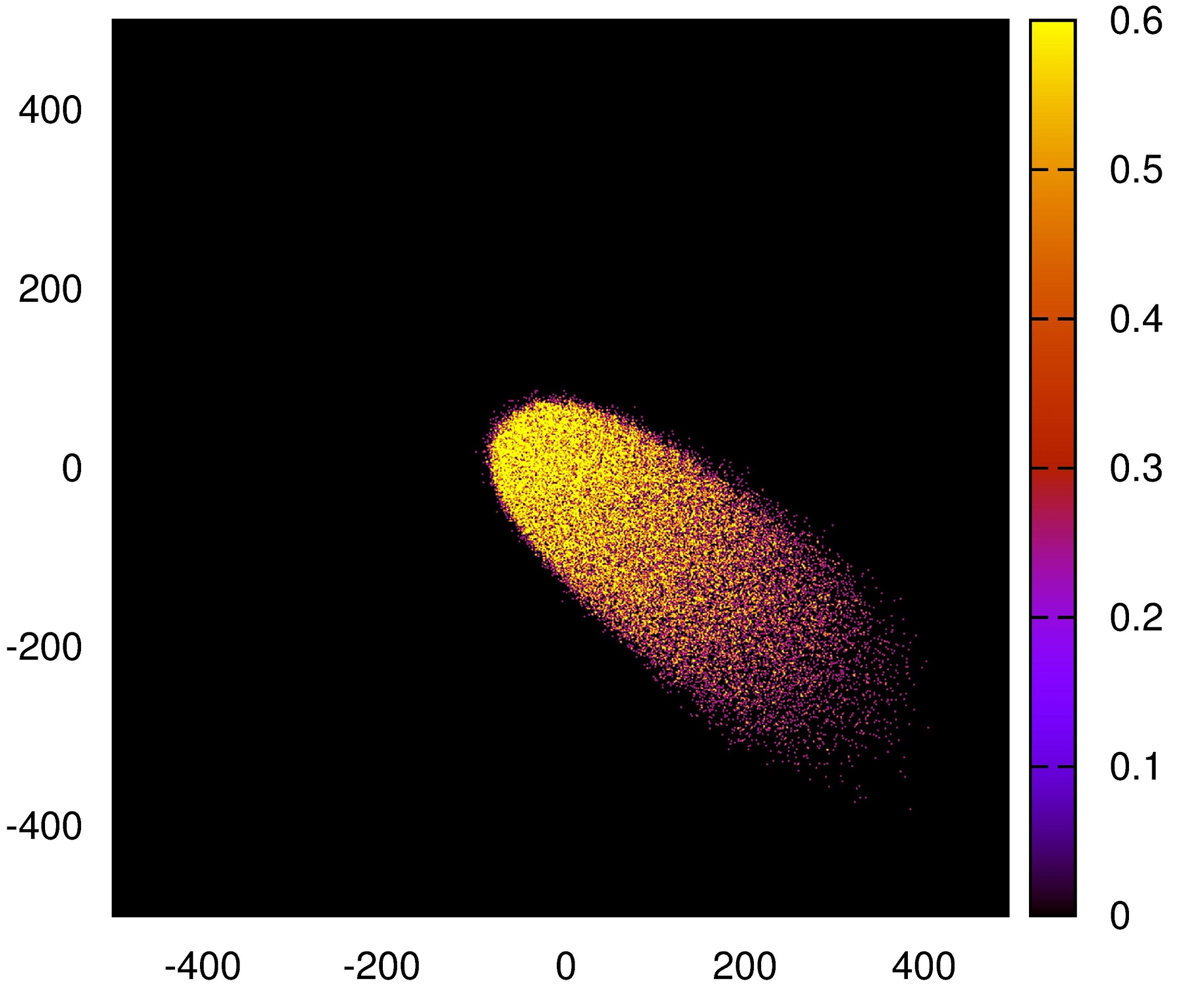}
\includegraphics[width=4.2cm]{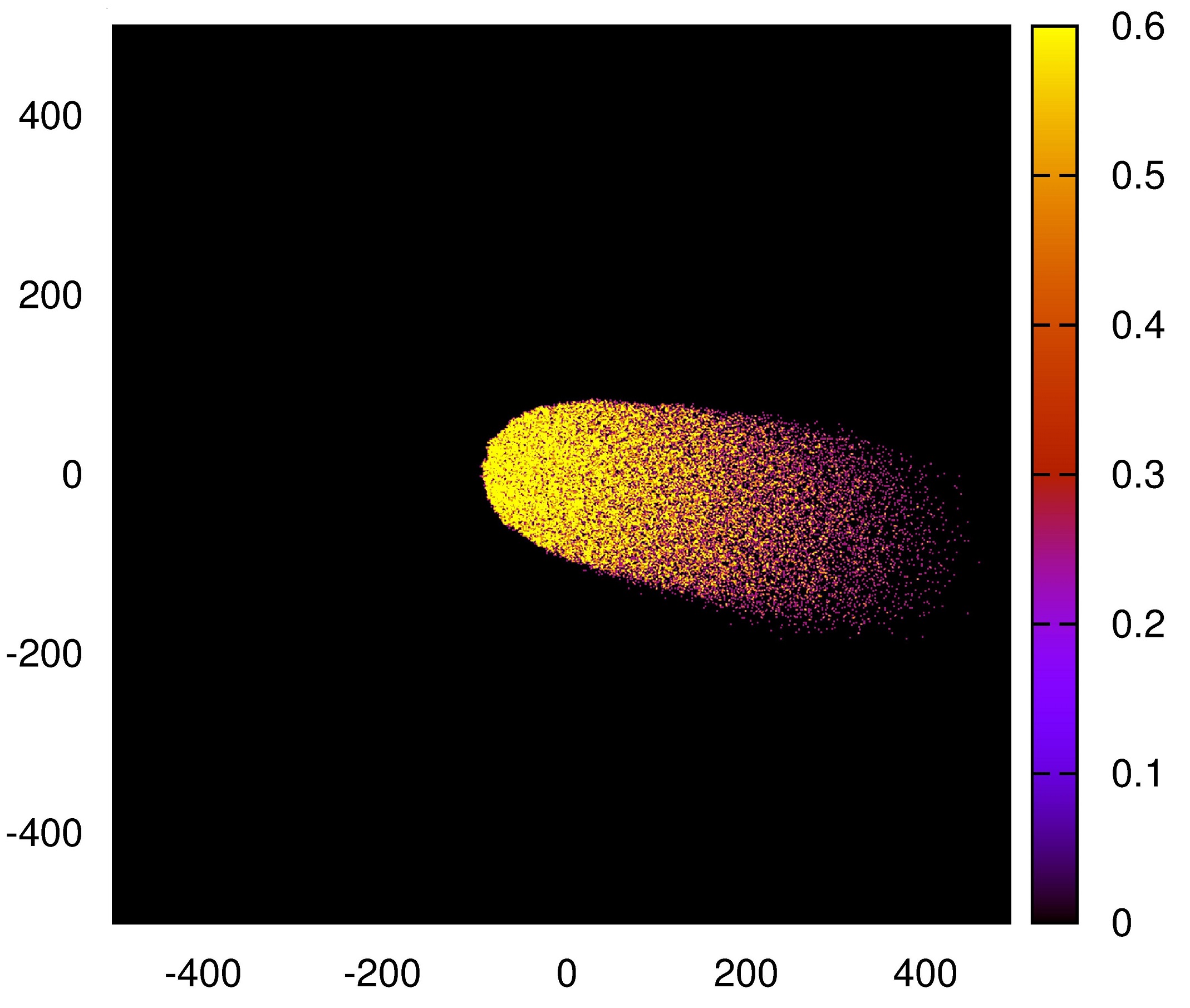}
\caption{Density plots of the results of the Monte Carlo simulations of diffusion in presence of an external force. At left hand side the driving force is $K=0.5$ and evolution time $t=700 \tau$ and at right  hand side plot is for  $K=7$ and $t=40 \tau$. The jump parameters are $W_x=3 a^2/\tau$, $W_y=0.1 a^2/\tau$, $V_x=0.1a^2/\tau$ and $V_y=2 a^2/\tau$.}
\label{bias}
\end{center}
\end{figure}
\section{Particles driven along anisotropic lattice}
\subsection{Single particle system driven along $x$ axis. }
When particles are driven along anisotropic lattice it is obvious that the direction of diffusion anisotropy and the direction of  driving  force compete.  Let us analyze the  situation, when particles are driven along   $x$ axis by arbitrary strong force, i.e. far above the region of the system linear response.  All transition rates in the positive direction of axis $x$ are multiplied by the  factor $e^{K/2}$ and all the transition rates in the reverse direction are multiplied by $e^{-K/2}$ \cite{kehr}.  Both factors describe the change of the jump rate values due to the  driving force $K$. Flux of driven particles can be calculated in  the case when they move independently of each other \cite{kehr}. In such a case the occupation  probability  of sites of different types are  $P_1=P_0 h_1$ and $P_2=P_0 h_2$, where $P_0$ is the equilibrium occupation probability for independent particles and $h_1$ and $h_2$ are factors modifying it in the presence of the force.
We calculate currents between two neighboring sites  \cite{kehr}
\begin{eqnarray}
J_{W_x}=W_x P_0 \left(e^{K/2}h_1-e^{-K/2}h_2\right) \nonumber\\
J_{V_x}=V_x P_0 \left(e^{K/2}h_2-e^{-K/2}h_1\right) \nonumber \\
J_{V_y}=V_y P_0 \left(h_1-h_2\right) \nonumber \\
J_{W_y}=W_y P_0 \left(h_2-h_1\right) \label{drive}
\end{eqnarray}
Total current in  $x$ direction is $J_x=0.5(J_{W_x}+J_{V_x})$ and in  $y$ direction $J_y=0.5(J_{W_y}+J_{V_y})$. We need to calculate $h_1$ and $h_2$. In the stationary state the sum of currents flowing in and out of each site is  equal to zero:
\begin{equation}
J_{W_x}+J_{V_y}-J_{V_x}-J_{W_y}=0
\end{equation}
Additionally due to conservation of the density we  have $0.5(h_1+h_2)=1$.   We obtain
\begin{eqnarray}
h_1=\frac{W_xe^{-K/2}+V_xe^{K/2}+W_y+V_y}{\left(W_x+V_x\right)\cosh{\left(\frac{K}{2}\right)}+W_y+V_y} \nonumber \\
h_2=\frac{W_xe^{K/2}+V_xe^{-K/2}+W_y+V_y}{\left(W_x+V_x\right)\cosh{\left(\frac{K}{2}\right)}+W_y+V_y}
\end{eqnarray}
\begin{figure}
\begin{center}
\includegraphics[width=4.2cm]{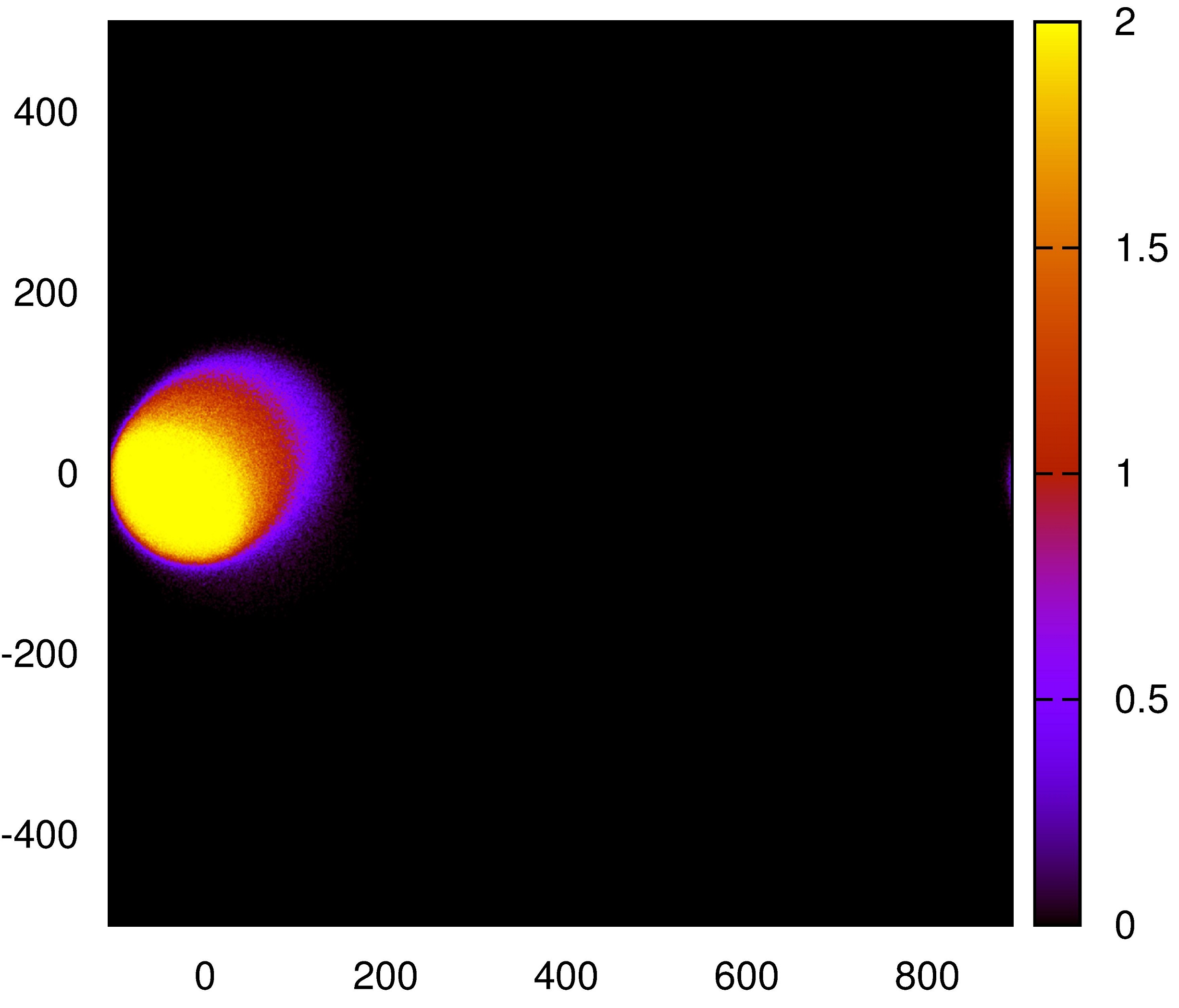}
\includegraphics[width=4.2cm]{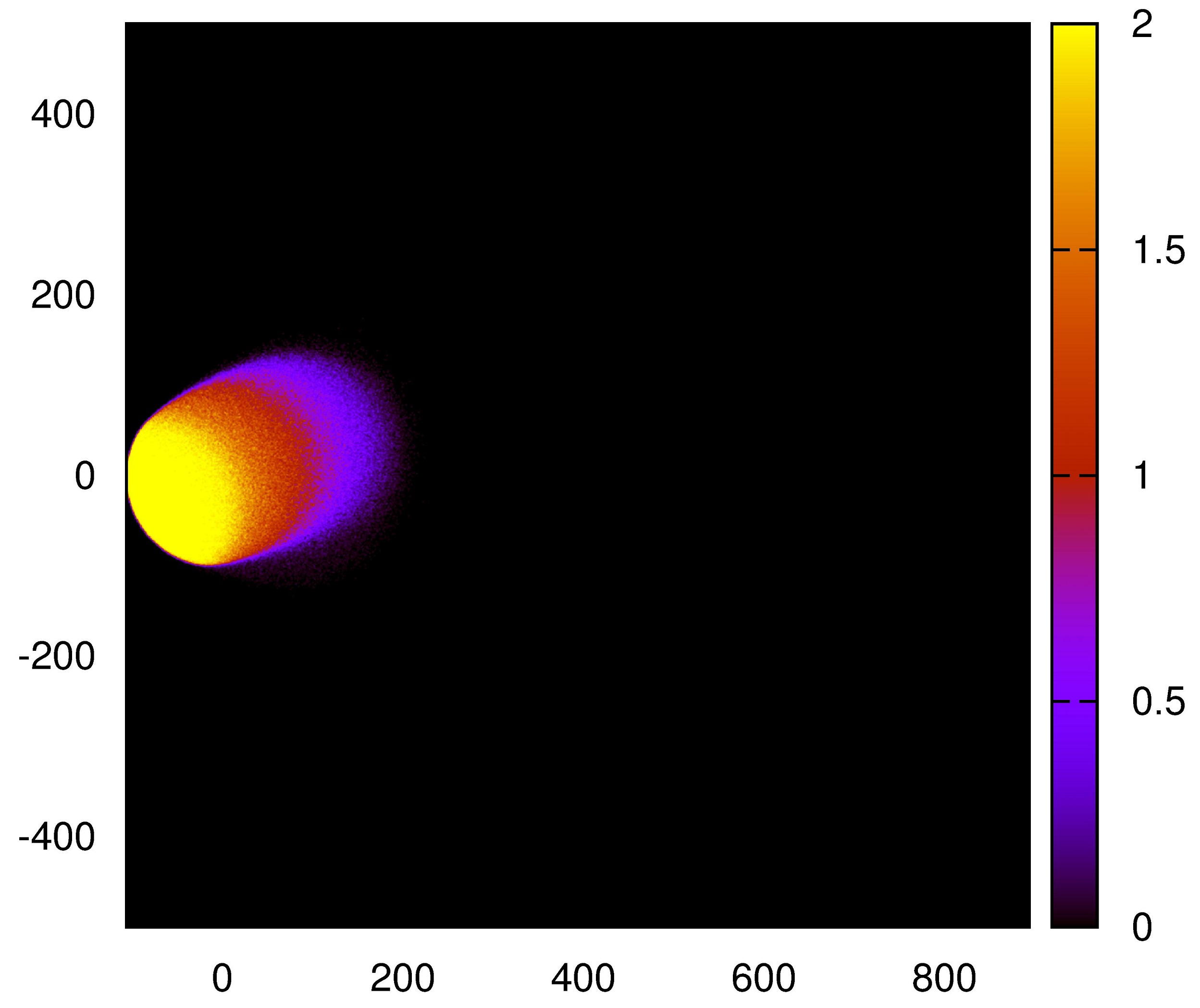}
\end{center}
\begin{center}
\includegraphics[width=4.2cm]{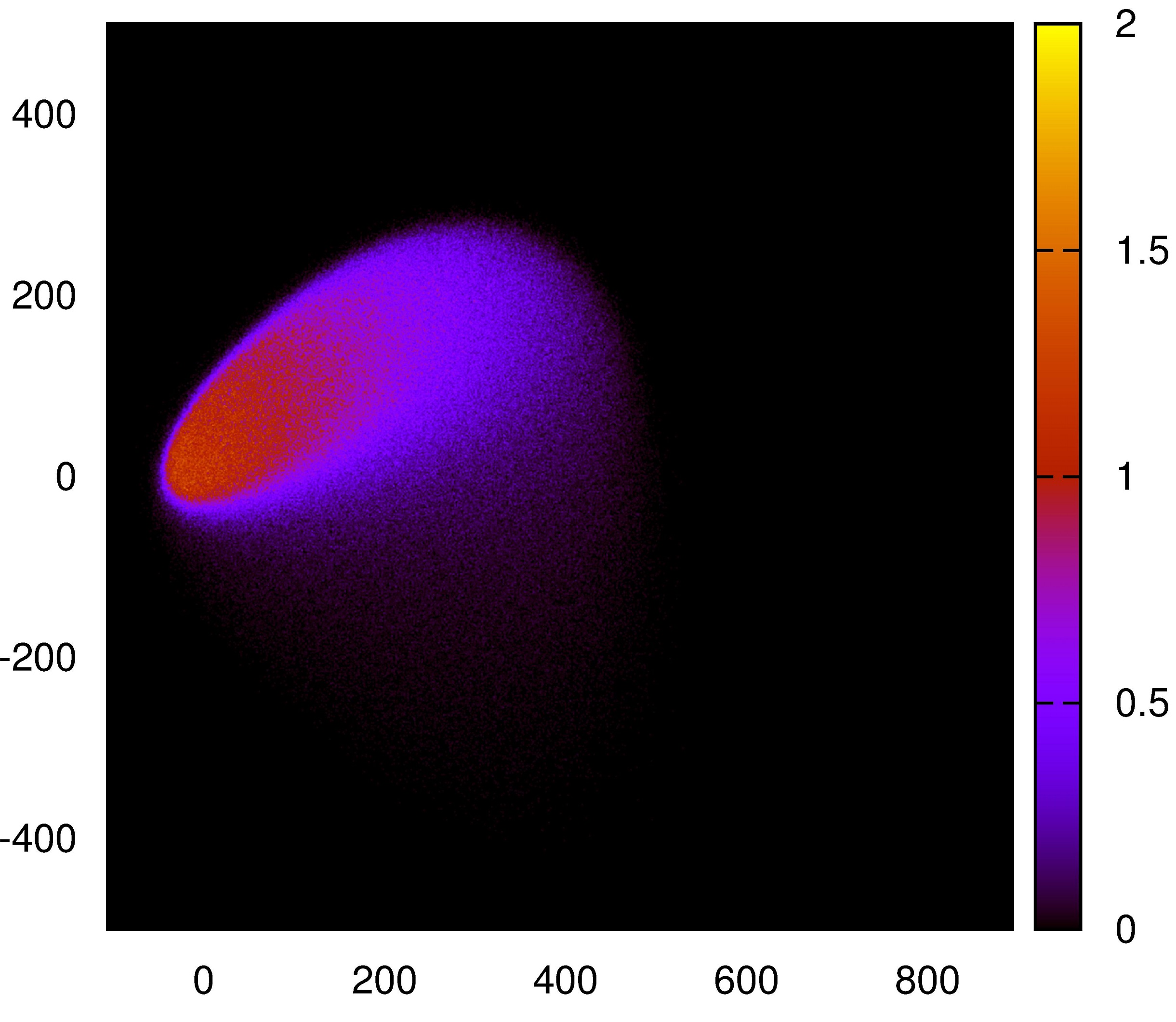}
\includegraphics[width=4.2cm]{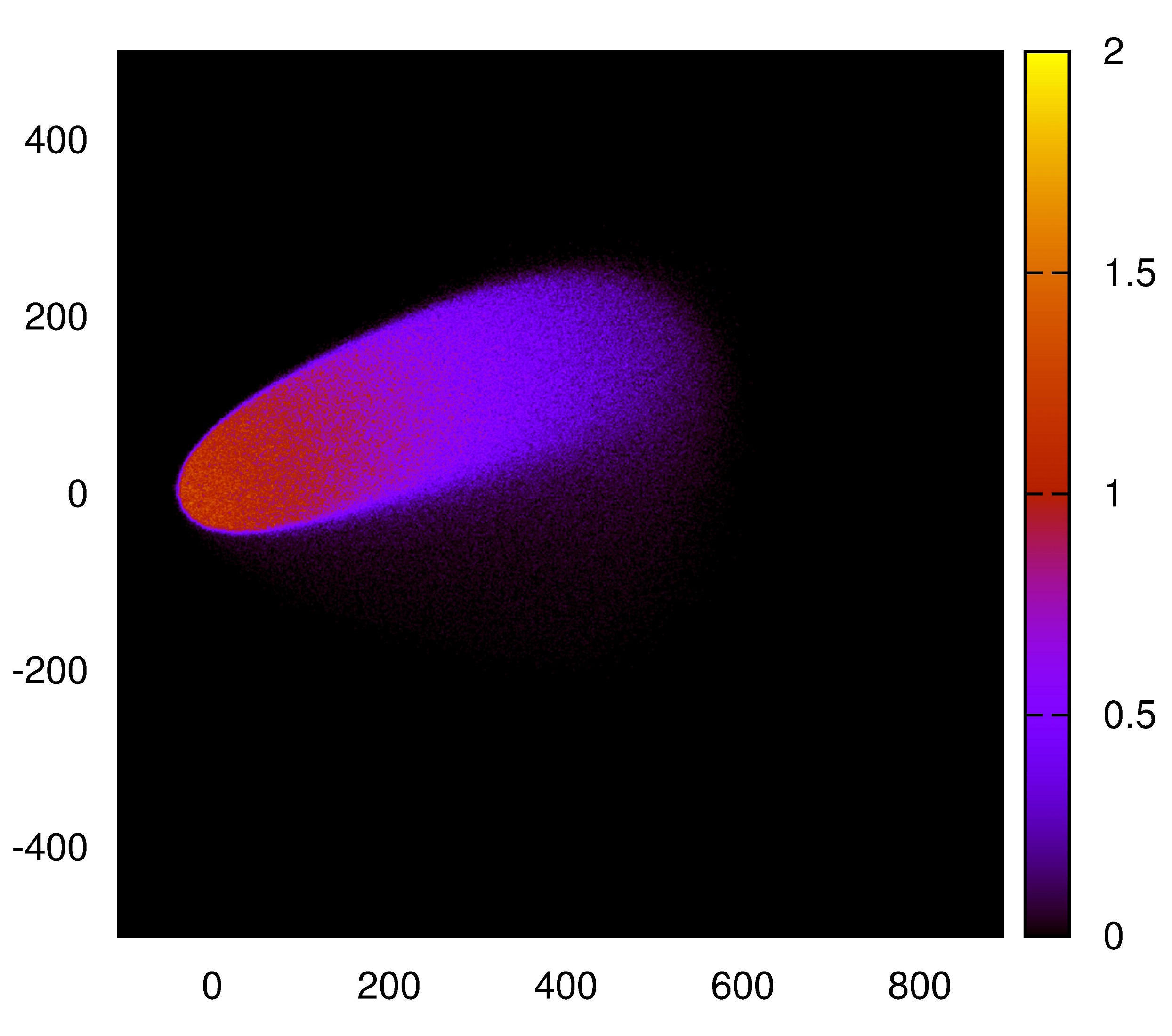}
\end{center}
\begin{center}
\includegraphics[width=4.2cm]{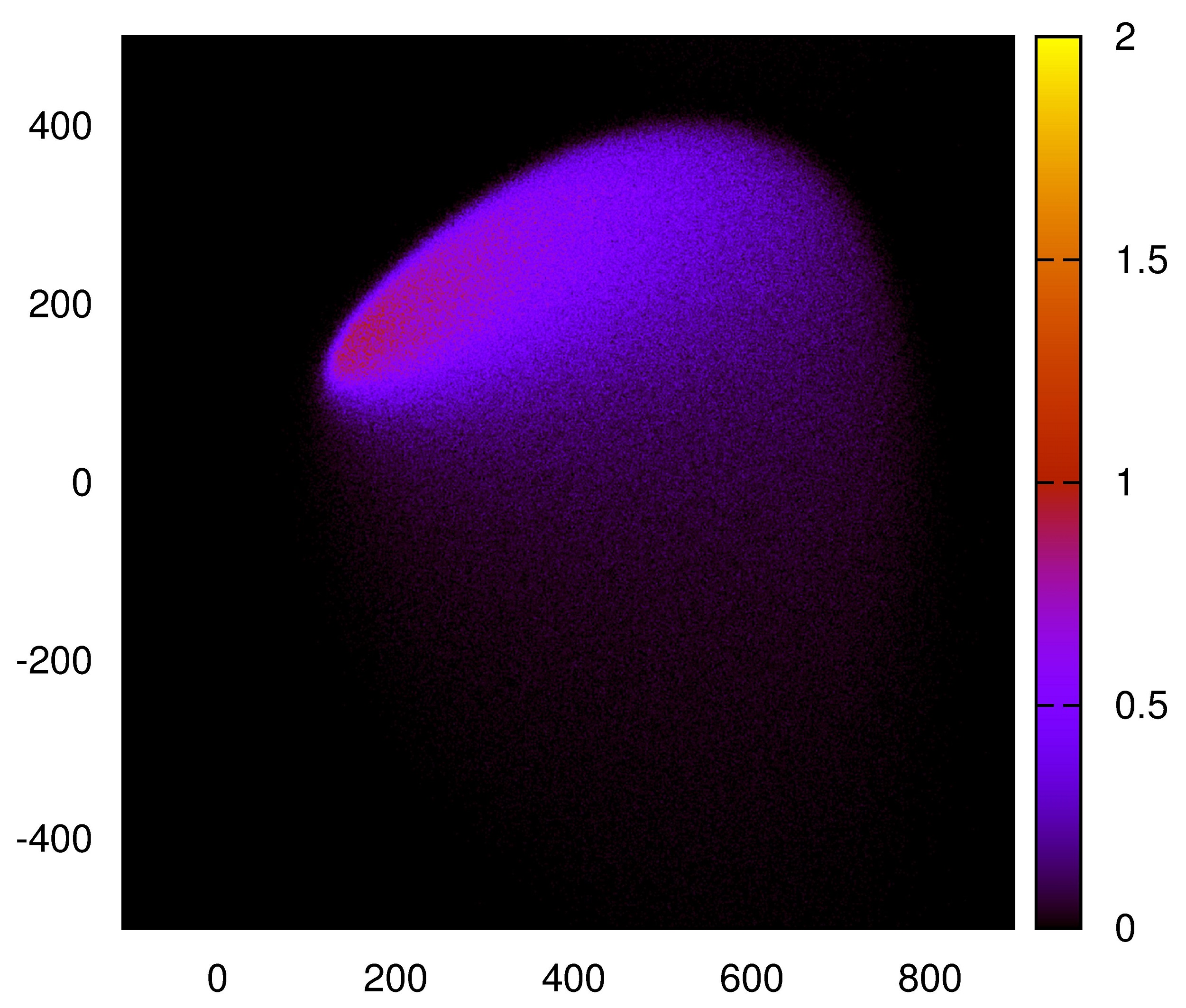}
\includegraphics[width=4.2cm]{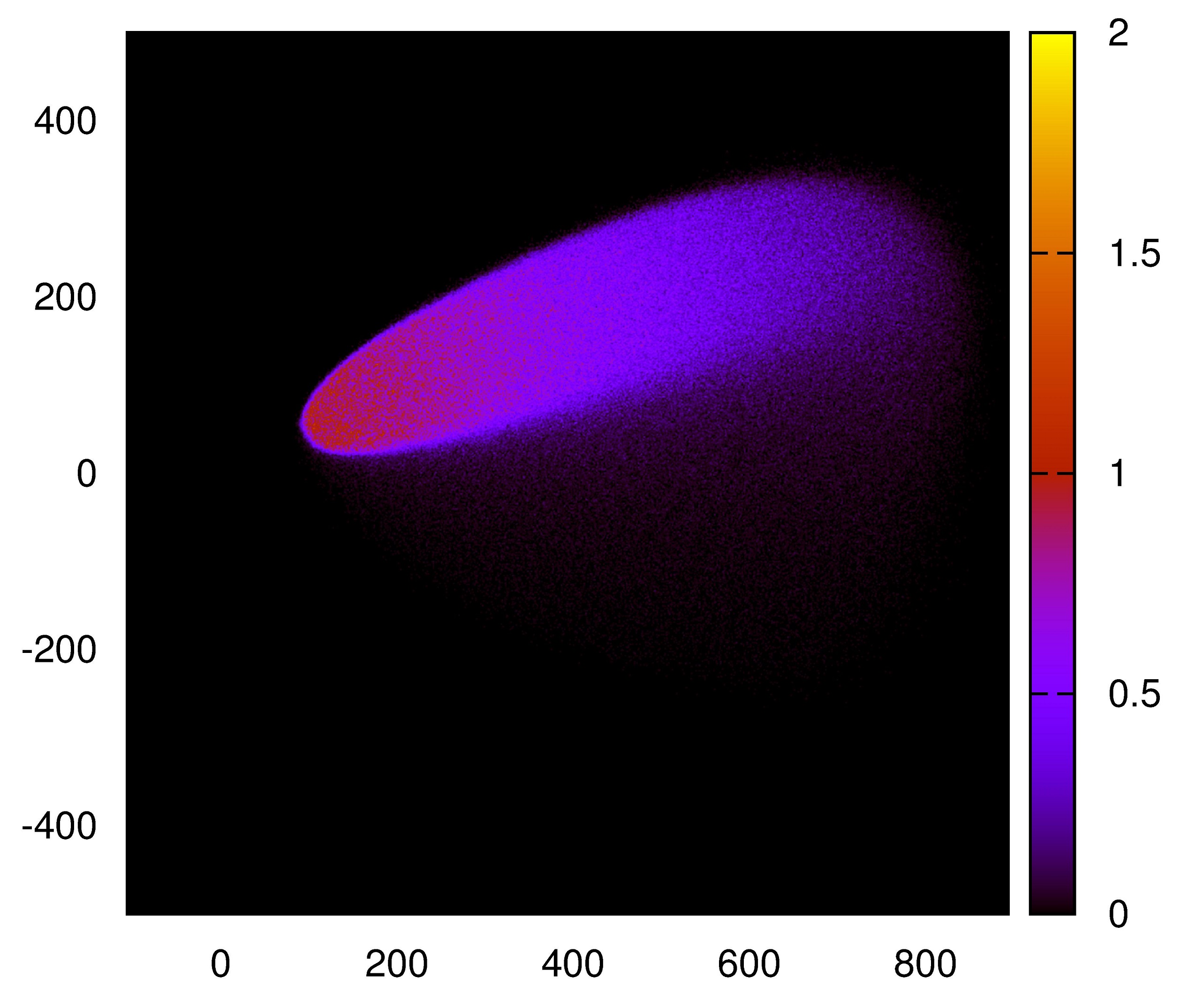}
\end{center}
\caption{Density plots of the results of the Monte Carlo simulations of driven mixture where individual particles diffuse with other jump rates than dimers. Jump rates are the same as in the Fig. \ref{interactions}.   Driving force is $K=0.5$ (left side) and $K=7$ (right side).  Coming from up down consecutive evolution stages are presented. At right hand side  $t=100 \tau, 800 \tau$ and $ 1500 \tau$ and at left hand side   $t=10 \tau, 50 \tau$ and $ 75 \tau$.  Jump parameters are the same as in  Fig. \ref{interactions}.}
\label{bias1}
\end{figure}
 Inserting this back to (\ref{drive}) we calculate the currents in a  system of particles of density $P_0=\theta$ for each direction   in the presence of the bias:
\begin{eqnarray}
J_x&=&\theta \frac{2W_xV_x\sinh{K}+\left(W_x+V_x\right)\left(W_y+V_y\right)\sinh{\frac{K}{2}}}{\left(W_x+V_x\right)\cosh{\frac{K}{2}}+W_y+V_y} \nonumber \\
J_y&=&\theta  \frac{-\left(V_x-W_x\right)\left(V_y-W_y\right)\sinh{\frac{K}{2}}}{\left(W_x+V_x\right)\cosh{\frac{K}{2}}+W_y+V_y}
\end{eqnarray}
We can also find the angle of the net current with respect to the $x$-axis
\begin{equation}
\phi=\arctan \left[ \frac{J_y}{J_x} \right] \label{angle_current}
\end{equation}
The dependence of  angle (\ref{angle_current}) on the driving force for the jump parameters  from Fig.\ref{no_interactions} is plotted by the solid line  in Fig.\ref{angle_bias}.  We see that for the  weak force $K=0$  particles move  close to the direction of diffusion anisotropy, which for the parameters used for the plot  is  equal to $-0.256 \pi$.    The direction of particle current approaches   the $x$ axis  for larger $K$ values from below  when particles are driven to the right  and from above  when they are driven to the left. Change of the flux direction from right down to left up happens at  $K=0$ and is seen  in Fig \ref{angle_bias} as  jump of a curve.  For  strong forces $K$  particles move closer to the  direction of driving force, along axis $x$.  This effect can be seen in the  MC simulation results   in Fig \ref{bias} where the evolution starting from the initial conditions shown in Fig \ref{initial} due to   two different forces $K=0.5$  and $K=7$  is compared. 

\subsection{Driven particle and dimer mixture.}
We can see now how the situation of the driven cloud of particles changes, when dimers move 
in other direction than independent particles do. We assume the same jump rates as studied before for the diffusing systems (Fig \ref{interactions}).  In Fig. \ref{angle_bias}  dashed curve  represents angle dependence on the driving force (\ref{angle_current}) in the case when the system contains only  dimers  i.e. with jump rates given by  $W\rq{}_x,W\rq{}_y, V\rq{}_x, V\rq{}_y$. It can be seen that driven dimers should move up whereas single particles move down while force acts along $x$ axis in its positive direction. The opposite situation happens when force is applied  in the negative  direction of the axis $x$.   We simulated  behavior of driven dense cloud of particles.  Initial state was the same as in  Fig. \ref{bias}. Results of the simulation are shown in Fig. \ref{bias1}. Cloud evolution  is  different from this for independent particles (Fig. \ref{bias}). Initial state consisted with dimers, so most of particles were driven  up jumping as dimers. They  have tendency to keep together rather than scatter  as it is visible in Fig \ref{bias}. The reason is that those particles that became separated  from dimers at the top of the cloud are forced to go down and they stay within whole group of particles forming dimers, while those separated from the bottom continue their motion down, so we can see diluted particle cloud below main part of the system. 

\begin{figure}
\begin{center}
\includegraphics[width=5cm]{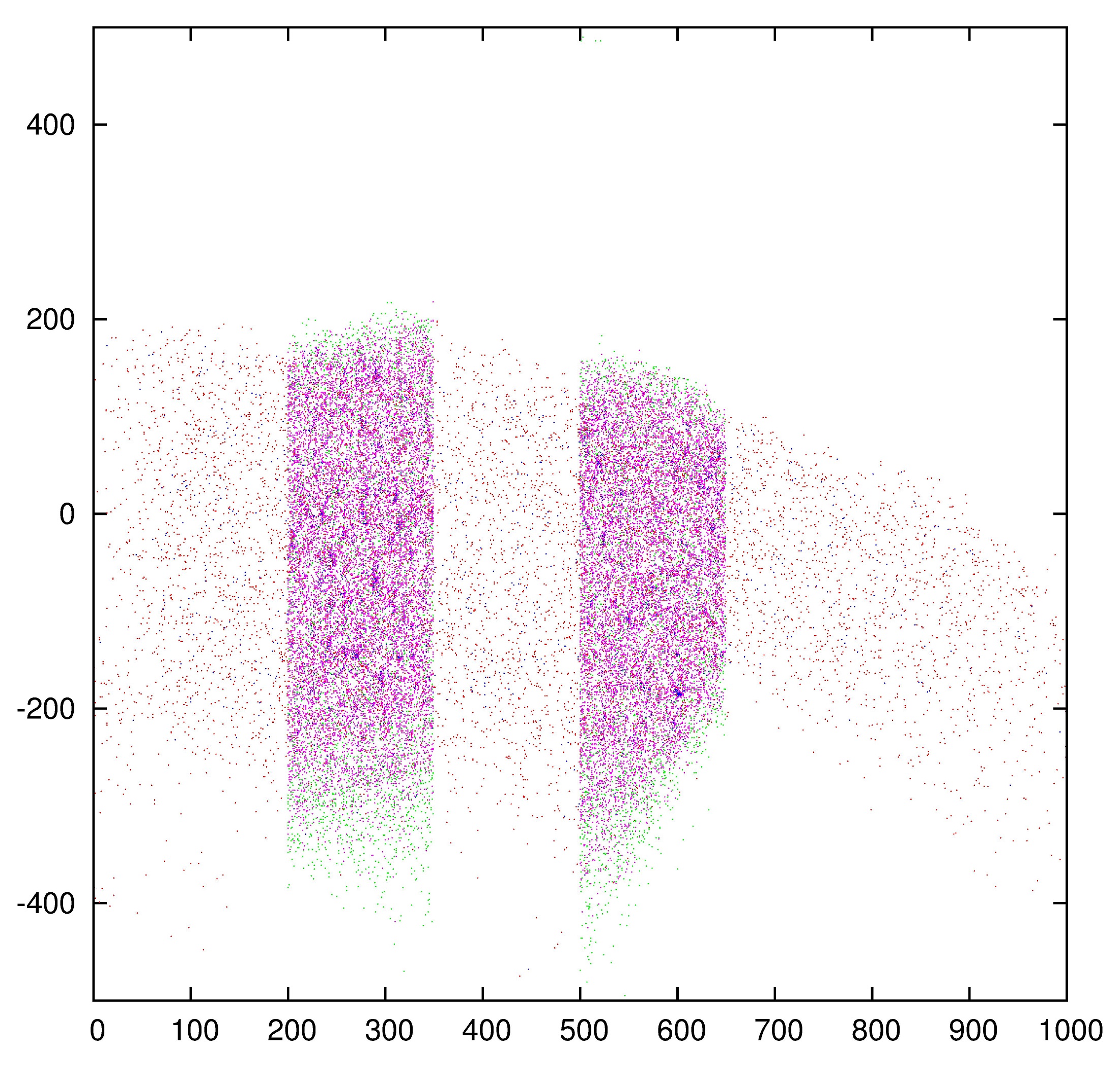}
\includegraphics[width=5cm]{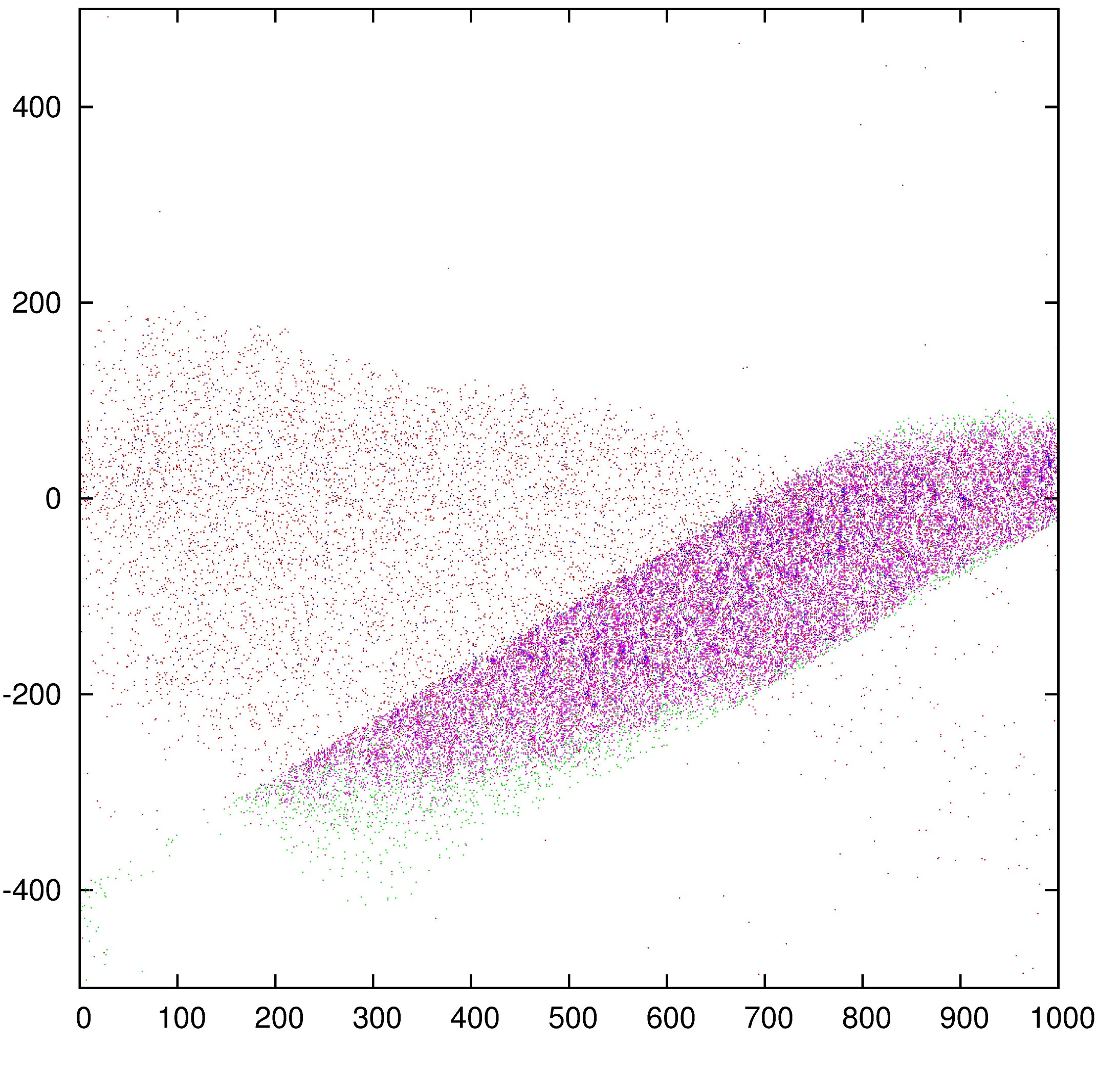}
\end{center}
\begin{center}
\includegraphics[width=5cm]{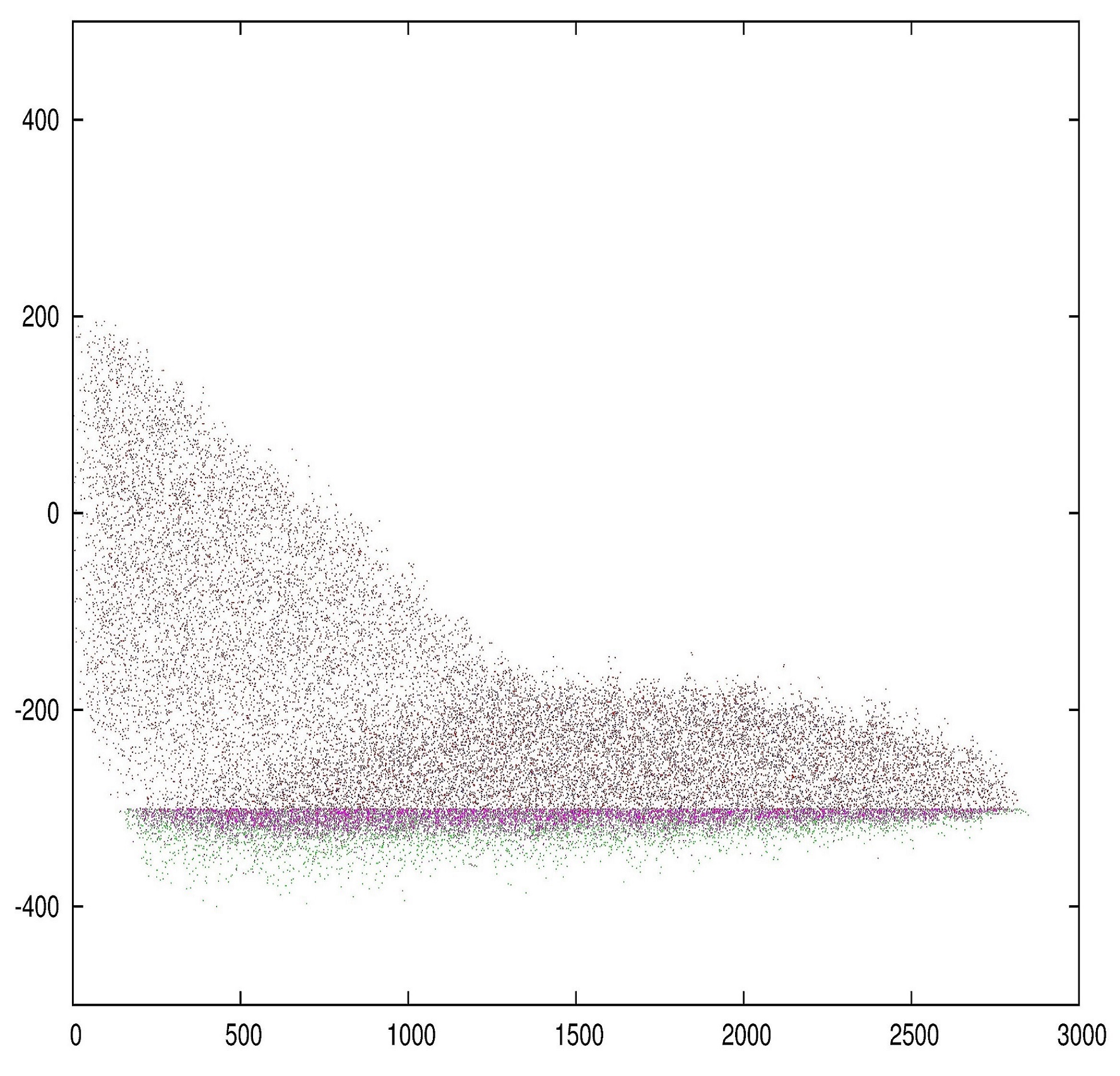}
\includegraphics[width=5cm]{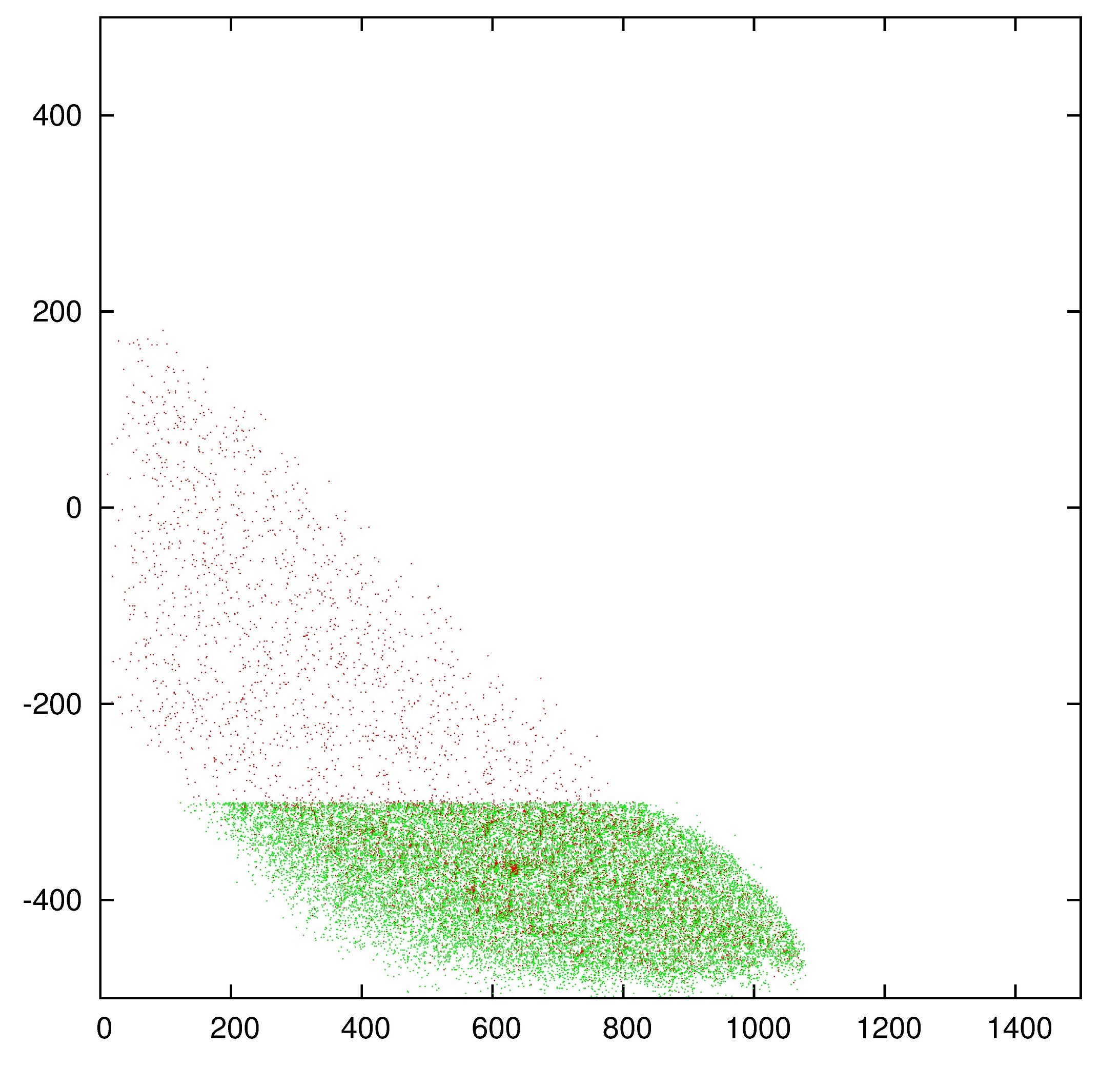}
\end{center}
\caption{Driven flux of particles over patterned surface.  Stripes with traps are oriented vertically at the top  left panel, at  30$^o$ to $x$ axis at right top,   horizontally it the left bottom  panel.  It the right bottom panel single particle system is shown coming across horizontal barrier with traps. Jump rates are the same as in Fig. \ref{interactions}. Jump rates out of trap sites are 0.001 and 0.2 fraction of sites are traps. Bias in all cases $K=1$. Green dots mark  particles in traps.}
\label{bias2}
\end{figure}
\section{Particle and dimer mixture at the defected surface areas}
Finally we show how the existence of trapping sites at the surface  changes direction of the moving particle cloud. Such trapping sites can be present at the surface due to different reasons:  defects, the existence of other particles or strain of the surface.
The mixture of  particles and dimers is  analyzed and as assumed above  particles move in different way than dimers.  The surface along which particles are driven contains some special regions with traps. Trapped particles can jump out from a  trapping site and probability of this process is assumed to be $0.001$. We study  flow of driven particle flux over surfaces with patterned  stripes  oriented in different ways. In these regions 0.2 fraction of sites have lower energy and work as trapping  sites. Their positions are chosen at random.
In Fig. \ref{bias2} we show how the system of driven particles behaves at the  patterned surface.  Particles enter the system on the left side between positions $-200 a$ to $200 a$ along $y$ axis building rather sparse cloud. They are forced to move to the right.  When they reach the patterned region  part of them are trapped and form dimers, which have different jump rates than individual particles. 
  Panel at the left top side  of the figure illustrates  particles coming through the vertical potential stripes. We can see that particles are moved up inside patterned stripes and they flow down in the region outside until they reach the second stripe with the same result. Next panel shows different orientation of patterned stripe. At the right top panel stripe is rotated 30$^o$ from $x$ axis. In such a case  particles also penetrate barrier with traps and, similarly as in the previous case, particles are driven up inside the stripe. When we rotate the stripe further  toward axis $x$, particles stop penetration of the barrier and start to slide along its border. It is very clearly seen in left bottom side  of Fig \ref{bias2}, where situation for the horizontal stripe is presented.  Here particles penetrate a short distance of the  region with traps, then next particles are reflected from this region.
The  behavior  of particles and dimers mixture  can be compared with the behavior of the system of single particles driven  across horizontal barrier in right bottom panel of Fig \ref{bias}. Particles  penetrate barrier without any change of the  direction of their motion. It can be seen that in the system of different   jump  anisotropy for particles and dimers properly oriented region of trapping sites can act as filter or particle mirror at the surface. It can lead to self-organization of the particle mixture.
\section{Summary}
Collective diffusion over anisotropic surfaces was analyzed. It was shown how the mixture of particles and dimers behaves when the  anisotropy of individual particle jumps is different than that for dimers. Jump anisotropy, defined by the relative values of consecutive jumps along $x$ and $y$ axes can be oriented in any direction, for particles and dimers independently.
Explicit expression for the diffusion coefficient matrix was derived for the single particle system and for the mixture. Diffusion in both cases was illustrated in the exemplary systems with almost perpendicular particle and dimer anisotropy axes. Driven diffusive  motion over the anisotropic surface  was studied in the region above the linear response limit. We have shown that when the easy diffusion axes of both mixture components are totally different the flow of diluted driven particle cloud changes its direction in the region of trapping centers at the surface.

\section{Acknowledgement}
Research supported  by the National Science Centre(NCN) of Poland
(Grant NCN No. 2011/01/B/ST3/00526). The authors would like to thank Dr. Z. W. Gortel for discussions and help in preparing the manuscript.


\end{document}